\documentclass[12pt]{elsarticle}
\usepackage{xcolor}

\usepackage{verbatim}

\usepackage[utf8]{inputenc}

\usepackage{graphicx}

\usepackage{amssymb}

\usepackage[modulo]{lineno}
\modulolinenumbers[2]

\usepackage[hyphens]{url}

\usepackage[bookmarks, breaklinks, hidelinks, pdfauthor={Weiming Hu}, unicode=true]{hyperref}

\usepackage{algorithm}
\usepackage[noend]{algpseudocode}

\hypersetup{linkcolor=blue,citecolor=blue,filecolor=black}

\usepackage[a4paper, total={6in, 9in}]{geometry}

\usepackage{outlines}

\usepackage{textcomp}
\usepackage{gensymb}

\usepackage{enumitem}

\usepackage{amssymb}

\usepackage{pifont}

\usepackage[numbers]{natbib}
\newlist{todo}{itemize}{2}
\setlist[todo]{label=$\square$}

\usepackage{caption}
\usepackage{subcaption}

\usepackage[toc, section=section, acronym]{glossaries}
\usepackage[toc, section=section]{glossaries}

\usepackage{glossary-mcols}

\setacronymstyle{long-short}

\setglossarystyle{mcolindex}

\newcommand*{\Fig}[1]{Figure~\ref{#1}}
\newcommand*{\Figs}[1]{Figures~\ref{#1}}
\newcommand*{\Table}[1]{Table~\ref{#1}}

\newcommand*{\Eq}[1]{Equation~(\ref{#1})}

\newcommand*{\Sect}[1]{Section~\ref{#1}}

\newcommand*{\App}[1]{\ref{#1}}

\usepackage{multirow}
\usepackage[normalem]{ulem}
\usepackage{adjustbox}
\useunder{\uline}{\ul}{}

\usepackage[figuresleft]{rotating}

\usepackage{listings}

\definecolor{codegreen}{rgb}{0,0.6,0}
\definecolor{codegray}{rgb}{0.5,0.5,0.5}
\definecolor{codepurple}{rgb}{0.58,0,0.82}
\definecolor{backcolour}{rgb}{0.95,0.95,0.92}

\lstdefinestyle{mystyle}{
  backgroundcolor=\color{backcolour},   commentstyle=\color{codegreen},
  keywordstyle=\color{magenta},
  numberstyle=\tiny\color{codegray},
  stringstyle=\color{codepurple},
  basicstyle=\ttfamily\footnotesize,
  breakatwhitespace=false,         
  breaklines=true,                 
  captionpos=b,                    
  keepspaces=true,                 
  numbers=left,                    
  numbersep=5pt,                  
  showspaces=false,                
  showstringspaces=false,
  showtabs=false,                  
  tabsize=2
}

\newacronym{ADAM}{ADAM}{Adaptive Moment Estimation}
\newacronym{AI}{AI}{Artificial Intelligence}
\newacronym{AnEn}{AnEn}{Analog Ensemble}
\newacronym{ANN}{ANN}{Artificial Neural Network}
\newacronym{API}{API}{Application Programming Interface}
\newacronym{ASOS}{ASOS}{Automated Surface Observing System}
\newacronym{ATSDR}{ATSDR}{Agency for Toxic Substances and Diseases Registry}
\newacronym{AVS}{AVS}{Agrivoltaic System}
\newacronym{Brier}{Brier}{Brier Score}
\newacronym{CDF}{CDF}{Cumulative Distribution Function}
\newacronym{CLI}{CLI}{Command Line Interface}
\newacronym{CMIP}{CMIP}{Coupled Model Intercomparison Project}
\newacronym{CONUS}{CONUS}{Continental United States}
\newacronym{CRM}{CRM}{Cloud Resolving Models}
\newacronym{CRMSE}{CRMSE}{Centered Root Mean Square Error}
\newacronym{CRPS}{CRPS}{Continuous Rank Probability Score}
\newacronym{CV}{CV}{Computer Vision}
\newacronym{DA}{DA}{Deep Analog}
\newacronym{EA}{EA}{Evolutionary Analog}
\newacronym{ECMWF}{ECMWF}{European Center for Medium Weather Forecasting}
\newacronym{ENSO}{ENSO}{El Ni\~no Southern Oscillation}
\newacronym{FLT}{FLT}{Forecast Lead Time}
\newacronym{GA}{GA}{Genetic Algorithm}
\newacronym{GCM}{GCM}{General Circulation Model}
\newacronym{GEFS}{GEFS}{Global Ensemble Forecast System}
\newacronym{GFS}{GFS}{Global Forecast System}
\newacronym{GHG}{GHG}{Greenhouse gas}
\newacronym{GISS}{GISS}{Goddard Institute for Space Studies}
\newacronym{GSI}{GSI}{Gridpoint Statistical Interpolation}
\newacronym{HF}{HF}{Heuristic Filter}
\newacronym{HPC}{HPC}{High-Performance Computing}
\newacronym{HRRR}{HRRR}{High-Resolution Rapid Refresh}
\newacronym{IDW}{IDW}{Inverse Distance Weighted}
\newacronym{IPCC}{IPCC}{International Panel on Climate Change}
\newacronym{IS}{IS}{Independent Search}
\newacronym{ITCZ}{ITCZ}{Intertropical Convergence Zone}
\newacronym{KF}{KF}{Kalman Filter}
\newacronym{LBR}{LBR}{Land-Based Renewables}
\newacronym{LOO}{LOO}{Leave-One-Out}
\newacronym{LSTM}{LSTM}{Long Short-Term Memory}
\newacronym{MAE}{MAE}{Mean Absolute Error}
\newacronym{ME}{ME}{Mean Error}
\newacronym{ML}{ML}{Machine Learning}
\newacronym{MOS}{MOS}{Model Output Statistics}
\newacronym{MRE}{MRE}{Missing Rate Error}
\newacronym{MSE}{MSE}{Mean Square Error}
\newacronym{NAM}{NAM}{North American Mesoscale Model}
\newacronym{NCAR}{NCAR}{National Center for Atmospheric Research}
\newacronym{NCEP}{NCEP}{National Centers for Environmental Prediction}
\newacronym{NOAA}{NOAA}{National Oceanic and Atmospheric Agency}
\newacronym{NWP}{NWP}{Numerical Weather Prediction}
\newacronym{OMT}{OMT}{Over Max Time}
\newacronym{PAN}{PAN}{Persistence Analog}
\newacronym{PAnEn}{PAnEn}{Parallel Analog Ensemble}
\newacronym{PDF}{PDF}{Probability Distribution Function}
\newacronym{PDO}{PDO}{Pacific Decadal Oscillation}
\newacronym{PEF}{PEF}{Parallel Ensemble Program}
\newacronym{PNA}{PNA}{Pacific/North American}
\newacronym{PV}{PV}{Photovoltaic}
\newacronym{PWS}{PWS}{Private Weather Station}
\newacronym{RAM}{RAM}{Random Access Memory}
\newacronym{RC}{RC}{Rank Correlation}
\newacronym{RCP}{RCP}{Representative Concentration Pathway}
\newacronym{RMSE}{RMSE}{Root Mean Square Error}
\newacronym{RPS}{RPS}{Rank Probability Score}
\newacronym{SAM}{SAM}{System Advisor Model}
\newacronym{SOM}{SOM}{Self-Organizing Map}
\newacronym{SS}{SS}{Schaake Shuffle}
\newacronym{SSE}{SSE}{Search Space Extension}
\newacronym{SSI}{SSI}{Spectral Statistical Interpolation}
\newacronym{SURFRAD}{SURFRAD}{Surface Radiation Budget}
\newacronym{SVI}{SVI}{Social Vulnerability Index}
\newacronym{SVM}{SVM}{Support Vector Machine}
\newacronym{TAU}{TAU}{Tuning and Analysis Utilities}
\newacronym{UHI}{UHI}{Urban Heat Island}
\newacronym{UMAP}{UMAP}{Uniform Manifold Approximation and Projection}
\newacronym{UML}{UML}{Unified Modeling Language}
\newacronym{USSE}{USSE}{Utility-Scale Solar Energy}
\newacronym{VGI}{VGI}{Volunteered Geographic Information}
\newacronym{WHO}{WHO}{World Health Organization}
\newacronym{WRCP}{WRCP}{World Climate Research Programme}
\newacronym{WRF}{WRF}{Weather Research and Forecasting}
\newacronym{NMM}{NMM}{Nonhydrostatic Mesoscale Model}
\newacronym{WU}{WU}{Weather Underground}
\newacronym{GHI}{GHI}{Global Horizontal Irradiance}
\newacronym{DNI}{DNI}{Direct Normal Irradiance}
\newacronym{DHI}{DHI}{Diffuse Horizontal Irradiance}
\newacronym{EW}{EW}{Equal Weighting}
\newacronym{RB}{RB}{Regime Based}
\newacronym{NN}{NN}{Nearest Neighbor}
\newacronym{IECC}{IECC}{International Energy Conservation Code}
\newacronym{USGS}{USGS}{U.S. Geological Survey}
\newacronym{EnTK}{EnTK}{Ensemble Toolkit}
\newacronym{MPI}{MPI}{Message Passing Interface}
\newacronym{STC}{STC}{Standard Test Condition}
\newacronym{STD}{STD}{Standard Deviation}
\newacronym{EIA}{EIA}{U.S. Energy Information Administration}
\newacronym{EBA}{EBA}{Ensemble-Based Application}
\newacronym{RCT}{RCT}{RADICAL Cybertool}
\newacronym{RS}{RS}{RADICAL-SAGA}
\newacronym{RP}{RP}{RADICAL-Pilot}

\title{A Scalable Solution for Running Ensemble Simulations for Photovoltaic Energy}

\author[1]{Weiming Hu}
\author[1]{Guido Cervone}
\author[2]{Matteo Turilli}
\author[2]{Andre Merzky}
\author[2]{Shantenu Jha}

\address[1]{Department of Geography and Institute of Computational and Data Sciences,\\Pennsylvania State University}
\address[2]{Department of Electrical and Computer Engineering, Rutgers University}

\begin{document}

\begin{abstract}
This chapter proposes and provides an in-depth discussion of a scalable solution for running ensemble simulation for solar energy production. Generating a forecast ensemble is computationally expensive. But with the help of Analog Ensemble, forecast ensembles can be generated with a single deterministic run of a weather forecast model. Weather ensembles are then used to simulate 11 10 KW photovoltaic solar power systems to study the simulation uncertainty under a wide range of panel configuration and weather conditions. This computational workflow has been deployed onto the NCAR supercomputer, Cheyenne, with more than 7,000 cores.

Results show that, spring and summer are typically associated with a larger simulation uncertainty. Optimizing the panel configuration based on their individual performance under changing weather conditions can improve the simulation accuracy by more than 12\%. This work also shows how panel configuration can be optimized based on geographic locations.
\end{abstract}

\maketitle

\section{Introduction}

The reliance on fossil fuels for power generation is not sustainable for meeting the increasing demand of a growing global economy and population. Building a sustainable society requires alternative sources for power production that can last for generations to come \cite{lewis2006powering}. Thanks to new incentives, regulations, improvements in technology, development of a competent workforce, paired with a general shift in sentiment against pollution and greenhouse emissions, renewable sources account for a significant portion of the overall energy production portfolio. 

In 2019, for the first time in history, more energy in the U.S. was produced from renewable sources than from coal \cite{eia2021}. Since early 2000s, the general trend of energy production in the U.S. showed a steady decline in coal, offset by an increase in natural gas, and renewable sources.  Predictions by the \gls{EIA} for 2050 suggest that 36\% of energy will be generated using natural gas (37\% in 2019), 38\% from renewables (19\% in 2019), 12\% nuclear (19\% in 2019), and only 13\% from coal (24\% in 2019) \cite{eia2020}. Therefore, while natural gas is predicted to remain almost constant over the next 30 years, renewable will double to offset coal and nuclear power.  Specifically, it is projected  that solar generation will account for almost 80\% of the increase in renewable generation through 2050 \cite{eia_dubin}.

Relying on solar power, however, can be challenging, despite the massive progress in utility-scale installation and the prevalence of residential \gls{PV} systems among communities. The sheer amount of available solar power does not necessarily guarantee its full utilization and integration into our electricity grid. This is due to rapidly changing weather conditions, astronomical factors, and other events that alter the amount of power produced.  In general, it can be said that there is a greater uncertainty associated with generating power using \gls{PV} because external weather conditions play a dominant effect. 

{\bf The key challenging problem is to match power production with demand, both of which are dynamic and variable. Furthermore, this must occur at different time-scales, ranging from day-ahead to seconds.} 

In the U.S. and most other countries, the energy production planning occurs in what is referred to as a day-ahead market \cite{ferruzzi2016optimal}. Each day, a portfolio of energy sources is created to meet a predicted demand.  A bidding process sets the price of electricity, where the different producers offer to sell electricity at a specific price.   Generally speaking, if a producer offers to sell a certain amount of energy at a price, not meeting the specific amount incurs in a penalty, and generating more energy leads to an opportunity loss \cite{wen2001strategic, davatgaran2018optimal}.   This process favors energy sources with smaller day-ahead uncertainty.

Traditional fossil fuel power sources, and some types of renewables such as hydroelectric power, have little day-ahead uncertainly, because the amount of potential power generation is directly proportional to the fuel available, may this be stored natural gas or volume of water in a reservoir.  On the other hand, \gls{PV} solar energy has larger uncertainty due changing weather conditions, astronomical factors, technological constraints, operational practices.

\textit{Weather:} \gls{NWP} are used to forecast day-ahead irradiance, which depends on general weather conditions \cite{alessandrini_analog_2015}.  While they are effective, they are also error prone, especially in case of rare or extreme events. For example, clouds \cite{chow2015cloud, kleissl2013recent} and precipitation are an important factor that changes the amount of solar irradiance reaching solar panels, but their day-ahead modeling is difficult, and often requires multiple parameterization schemes. Similarly, aerosols interact with the incoming solar irradiance \cite{wan_photovoltaic_2015, jimenez_wrf-solar_2016} by increasing the diffused component of insolation. Any changes in irradiance, which can either decrease or sometimes increase, leads to an additional uncertainty in power production.  Aerosols are present in larger quantities and thus have stronger impact in areas with heavy air pollution, and transient events like desert storms or volcanic eruptions.  Because of the uncertainty in weather forecasts, it is currently not possible to always and consistently forecast day-ahead \gls{PV} solar energy production.  

\textit{Astronomical:} Additionally, \gls{PV} solar energy is not available during the night, and daytime production changes throughout the year depending on sunrise and sunset times, and solar elevation.  While this variability can be easily forecasted and accounted for, it still requires a daily optimization to properly estimate future energy production.

\textit{Technological:} Another source of uncertainly is intrinsically related with the electro-chemical processes that \gls{PV} panels rely upon to generate electricity. The efficiency of power generation varies based, in addition to irradiance, also on cell temperature \cite{dirnberger_impact_2015, huld_estimating_2015}, which itself is related to ambient temperature. Wind speed affects air temperature, and thus must be taken into account to make accurate predictions \cite{mills_implications_2011, al-dahidi_analysis_2020}.   Once irradiance and temperature are forecasted, the conversion to power generation can be made through a \gls{PV} panel simulator. However, the nominal panel power capacity is evaluated under \gls{STC}, which might be different than real-world performance. For example, the \gls{STC} specifies a cell temperature of 25°C and an irradiance of 1000 $W/m^2$ with an air mass 1.5 spectrum. These conditions are only rarely if ever met.  

\textit{Operational:} Finally, panels must be maintained clean and they generally degrade with time, leading to their efficiency changing constantly.  Given two identical weather conditions, the same panel might generate a different amount of electricity at different times.

\subsection{Solar Power Forecasting}

Solar power forecasting can be viewed as a two-step problem: characterizing the weather and the \gls{PV} panels.  There are many different methods to solve these two problems, sequentially or concurrently, and they depend on factors such as the temporal and spatial scale of the results, the temporal resolution, if a measurable level of uncertainty is required, and the computational resources available.  

Solar power forecasting is usually divided based on the temporal forecasting horizon \cite{gensler_analog_2016}. Forecasting up to six hours is referred to as short-term, and it is where extrapolation and persistence methods excel due to their computational efficiency and good accuracy \cite{pedro_assessment_2012}. Following, is mid-term forecasting, spanning from six to 72 hours, where physical and/or statistical numerical modeling is considered the most effective technique. Because of the rapidly changing environmental conditions, the immediate ambient history of is not a good predictor for future outcomes. 

For day-ahead mid-term forecasts which are the focus of this research, the weather part is solved using \gls{NWP} models, that resolve weather features and forecast future conditions (e.g., cloud cover, wind speed, temperature). They are run at different temporal and spatial scale, and can be optimized for specific locations and weather patterns. However, weather forecasts do not directly translate into amount of power generated, and specific characteristics of the \gls{PV} array and its installation, as well as operational status, must be taken into account.  

The second step can be solved using solar panel numerical simulators or simple empirical functions; or using statistical and computational methods based on past measurements.  Which method to use is dictated by the specific needs and the level of accuracy required, but in general it can be said that simulators are less effective at accounting for the specific aspects of installation and location, while statistical and computational solutions are limited to historical patterns which might under-represent conditions, most importantly, rare or extreme events. The latter methods can also only be used in the presence of a robust history of measurements, and thus are generally unsuitable for feasibility studies. The results are also very location dependent, and while they might be optimal at predicting a specific panel configuration and installation, their results are likely to be so specific that cannot be generalized to other locations or installations.   Finally, because panels degrade over time, a statistical correction is also needed to capture this decrease in efficiency.

Among the statistical and computational methods, \gls{ANN}  have been used to predict both the behavior of panels, and in some cases also the weather conditions. A plethora of literature \cite{al-dahidi_analysis_2020, pedro_assessment_2012, chen_online_2011, omar_day-ahead_2016, sperati_application_2016, nitisanon_solar_2017, cervone_short-term_2017, chen_solar_2017, kumar_artificial_2018, theocharides_day-ahead_2020} have emerged to study the application of \gls{ANN} on solar power forecasting because they can be trained to focus on a geographically confined region and optimize local solar power forecasting. The general idea is to train an \gls{ANN} to learn the statistical pattern between weather variables and the observed power production. This technique usually requires a historical archive of weather observations together with the associated power production records. A properly trained \gls{ANN} is able to model the condition of solar panels and the surrounding environment like shadowing.

\subsection{Quantification of Uncertainty}

The simulation of a power production system is a highly complex and volatile process. Uncertainty arises when a complete and exact information on a process is lacking.

To simulate the amount of power generated, information regarding the weather (solar irradiance and temperature) and the panel (installation and condition) are needed. However, both are subject to a lack of information, and therefore, uncertainty. The atmosphere is a chaotic system that cannot be precisely observed and as a result, ``the prediction of the sufficiently distant future is impossible by any means'' \cite{lorenz1963}. Observations often have limited spatio-temporal resolution and  \gls{NWP} models only approximate the true physical processes and atmospheric interactions.

With regards to the simulation of panel performance, uncertainty can originate from an imprecise representation of the surrounding environment, e.g. shadowing, and from an incomplete characterization of real world performance  under a wide range of possible weather conditions, since the panels are typically tested under the \gls{STC}.

Simulation ensembles are particular helpful in quantifying uncertainty in the forecasts. In weather forecasting, \gls{NWP} models can be run many times to generate a range of possible deterministic realizations of the future state of the atmosphere. This ensemble of possible future states provides a quantifiable representation of the uncertainty for a given forecast. If ensemble members are very different, referred to as  ensemble spread, the specific uncertainty for the forecast is high.

There are many approaches for constructing a forecast ensemble, for example, applying varying perturbations applied to initial state variables (i.e. wind speed, temperature), changing dynamics schemes or parameterizations of an \gls{NWP} model, or using stochastic means for perturbing physical parameterizations \citep{berner2009spectral, haupt2017principles}. A hybrid approach can also be devised by combining one or more of the aforementioned methods. However, these approaches involve running \gls{NWP} models multiple times, which is computationally expensive.

The analog-based approach, \gls{AnEn}, has been proposed \cite{delle2013probabilistic} to construct forecast ensemble without extra runs of \gls{NWP} models. The \gls{AnEn} is a technique to generate forecast ensembles with a deterministic \gls{NWP} model and the corresponding historical observations. It does not require extra runs of the deterministic model which already offers a great reduction in the computational cost of ensemble generation. It has been successfully applied to a series of forecasting problems, including surface variables \cite{delle2013probabilistic, junk2015analog, hu2019dynamically, alessandrini2019improving} and renewable energy production \cite{vanvyve2015wind, alessandrini_analog_2015, junk2015predictor, cervone_short-term_2017, alessandrini2020schaake}. An in-depth introduction on the \gls{AnEn} is provided in \Sect{sect:MultiAnEn}.

\subsection{Scope and Organization}
This chapter introduces a scalable workflow that focuses on evaluating solar power production and its predictability from geographically diverse regions. Instead of relying on ground observations, a \gls{NWP} model is coupled with a \gls{PV} solar power simulation system which tests the predictability of 11 different \gls{PV} modules. These \gls{PV} modules vary in many aspects including material, size, and efficiency.  

Power forecasts are generated by using an \gls{AnEn} ensemble of \gls{NWP} forecasts that characterize future weather conditions, where each member of the ensemble is used as input for the panel simulator.  Therefore, for any given day 21 weather conditions are each used to generate 11 power predictions, where 21 is the size of the weather ensemble and 11 is the number of the panel configurations.   This is repeated for every grid over \gls{CONUS} (~60,000) every hour for the daylight time of the day for 2019, leading to over 60 trillion computations.  The system is verified by using the analysis fields of the weather model, which is the closest complete and high resolution geographical estimation of real-world weather conditions.  

Results show that \gls{PV} modules have various levels of predictability and they should be chosen based on optimizing both the power production and the predictability to support solar energy penetration. This workflow is designed to be scalable because of the very large number of computations required, which can scale significantly if testing more panel configurations, or increasing further the spatial and temporal scope or their resolutions.

The rest of the chapter is organized as follows: \Sect{sect:data} introduces the research data used in this project; \Sect{sect:method} describes the main methodology of generate ensemble forecasts with the \gls{PAnEn} implementation \cite{huparallel} and its integration with a power simulation system implemented with \textit{PV\_LIB} \cite{holmgren_open_2016}; \Sect{sect:results} shows results for solar power simulation with a total of 11 \gls{PV} modules; \Sect{sect:entk} specifically discusses the design of the scalable workflow with the RADICAL \gls{EnTK} \cite{balasubramanian2016ensemble}; and finally, \Sect{sect:final} summarizes the chapter.

\section{Research Data}
\label{sect:data}

\subsection{The North American Mesoscale Forecast System}

The \gls{NAM} is a weather forecast model operated by the \gls{NCEP}. It aims to provide short-term deterministic forecasts for weather variable at a variety of vertical levels and on multiple nested domains. Currently, the \gls{WRF} \gls{NMM} is run as the core of \gls{NAM} after its replace of the Eta model in 2006, and since then, the three names are interchangeable, \gls{NAM}, \gls{WRF}, and \gls{NMM}, typically referring to the same model output.

\gls{NAM} is initialized four times per day at 00, 06, 12, and 18 UTC. It is run with multiple nested domains with differences in the geographic coverage and the spatio-temporal resolution. In this study, the parent domain (\gls{NAM}-\gls{NMM}) with a 12 km spatial resolution is used. This domain covers North America. The production run provides forecasts up to 84 hours into the future, namely an 84-hour \gls{FLT}, with the first 36 hours being hourly and the rest of the lead times being every 3 hours. The parent domain run is initialized with a 6-hour \gls{DA} cycle and it is updated hourly using the hybrid variational ensemble \gls{GSI} and the \gls{NCEP} Global Ensemble Kalman Filter method. The analysis production is usually referred to as \gls{NAM}-ANL. There are other one-way nested domains within the parent 12 km domain, e.g., \gls{CONUS}, Alaska, Hawaii, and Puerto Rico. These nested domains have a fixed 3 km spatial resolution and a 60-hour \gls{FLT}. However, only the parent grid products have a historical data archive since 2006 available from \gls{NCEP}. 

In this study, \gls{NAM}-\gls{NMM} initialized at 00Z and \gls{NAM}-ANL initialized at 00, 06, 12, and 18 UTC, are collected between 2017 and 2019. Since the day-ahead \gls{PV} energy market is of particular interest and \gls{CONUS} covers four time zones, the \gls{FLT}s from 06Z to 32Z from \gls{NAM}-\gls{NMM} are focused on throughout the study. This lead time period covers from 01:00 (the day of initialization) to 03:00 (the next day) in the Easter time zone, and from 22:00 (the day prior to initialization date) to 00:00 (the next day) in the Pacific time zone. There is no compensation for day-light saving when calculating the local time to simplify the computation and analysis. \gls{NAM} provides a few hundred variables. \Table{tab:Variables} summaries the variables used in this study.

\begin{table}
\caption{Variables from \gls{NAM}-\gls{NMM} and \gls{NAM}-ANL used in this study}
\centerline{
\scalebox{0.9}{
\begin{tabular}{llll}
\hline\hline
Variable & Short name & Unit & Vertical level \\ \hline
Downward short-wave radiation flux & dswrf & $W/m^2$ & Surface \\
Albedo & al & $\%$ & Surface \\
Pressure & sp & $Pa$ & Surface \\
Orography & orog & $m$ & Surface \\
Temperature & 2t & $K$ & 2 m above ground \\
U wind component & 10u & $m/s$ & 10 m above ground \\
V wind component & 10v & $m/s$ & 10 m above ground \\
Total Cloud Cover & tcc & $\%$ & Atmosphere as a single layer \\
\hline\hline
\end{tabular}}}
\label{tab:Variables}
\end{table}

\begin{figure}
    \centering
    \includegraphics[width=\textwidth]{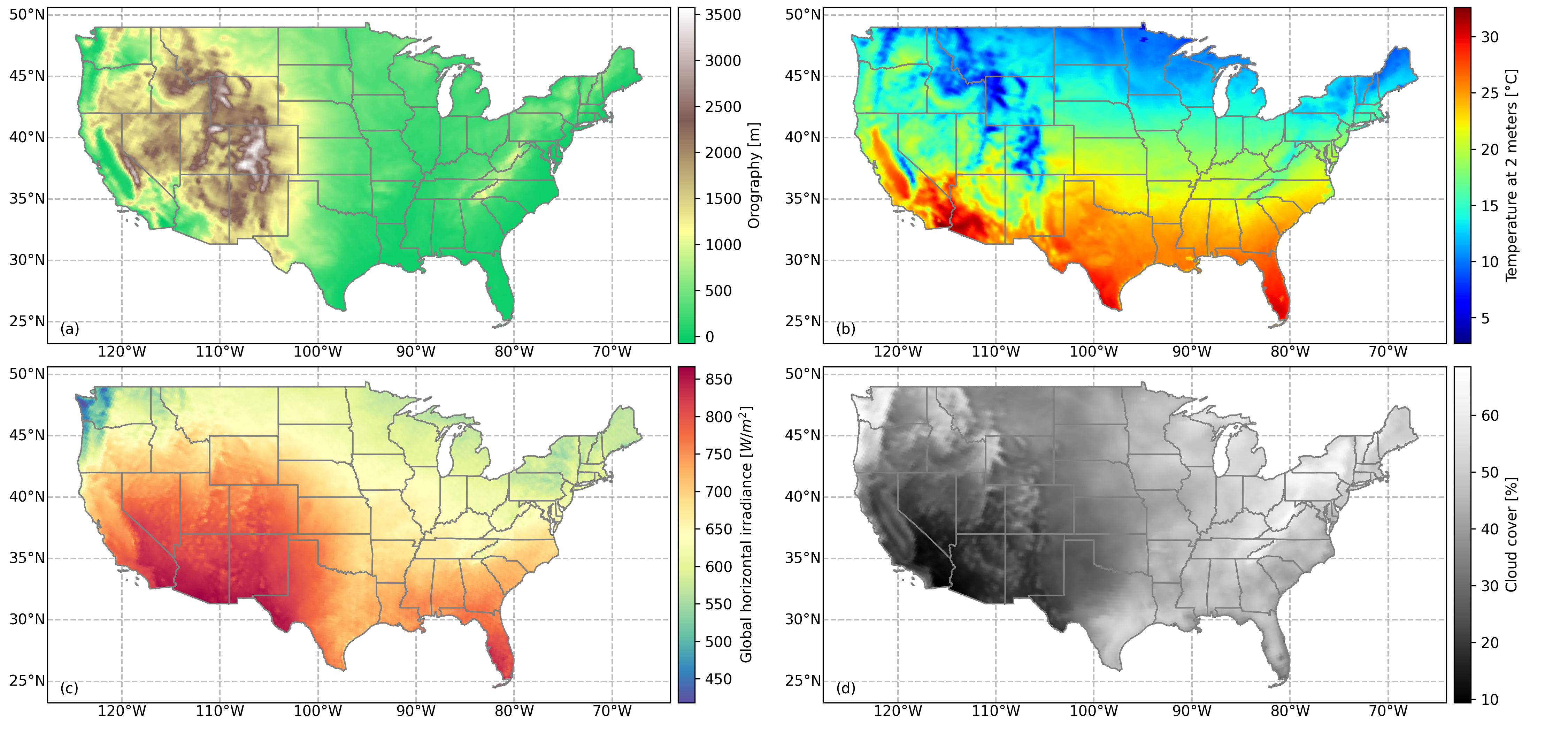}
    \caption{\gls{NAM}-ANL over \gls{CONUS} for 2018: model orography (a), average daily maximum temperature at 2 meters (b), average daily maximum \gls{GHI} (c), and average cloud cover (d). State boundaries are shown in solid lines.}
    \label{fig:Analysis2018}
\end{figure}

\Fig{fig:Analysis2018} shows the \gls{NAM} orography in (a) and the annual average of daily maximum temperature, daily maximum \gls{GHI}, and cloud cover respectively in (b, c, d) for 2018. The terrain representation of \gls{NAM} is aggregated from a high resolution \gls{USGS} 30' elevation dataset. The horizontal resolution of the \gls{USGS} 30' dataset is about 0.6 km at 40°N. The elevation of the center of each individual \gls{NAM} grid cell is set to the average elevation of the area in the \gls{USGS} dataset. A smoothing operation has been originally carried out to reduce the level of noise in the high resolution elevation dataset. As shown in \Fig{fig:Analysis2018}a, the model orography reflects well the elevation variation at and west to Rocky Mountains and at Appalachian Mountains. The spatial features of Coastal Ranges are also legible from the map. \gls{CONUS} has complex terrain features with elevation ranging from -74 m to 3577 m. \Figs{fig:Analysis2018}b, c show the annual average of daily maximum temperature and \gls{GHI} respectively. Daily maximum, instead of average, are calculated from the hourly analysis field. This is to avoid skewing the statistics by including night times of which are not of particular focus. Surface temperature exhibits typical longitudinal patterns with exceptions to region regions with drastic topographic changes. \gls{GHI} has a strong correlation with temperature, exhibiting higher levels at Rocky Mountains, the Great Basin, and Coastal Plains near Florida, compared to the other regions. Adversely, Coast Ranges at west Washington and west Oregon has a low level of annual \gls{GHI}. This can be explained by the annual average of cloud cover, as shown in \Fig{fig:Analysis2018}. Significant cloud cover is observed near these regions, blocking the incoming solar radiation. Cloud cover is one of the governing parameter for \gls{GHI}, and therefore, they are highly correlated. In terms of \gls{PV} power production, however, ambient temperature also plays an important factor because it affects the temperature of \gls{PV} panel module cells which, in turn, affects the efficiency of power generation.

\subsection{IECC Climate Zones}

\begin{figure}
    \centering
    \includegraphics[width=0.8\textwidth]{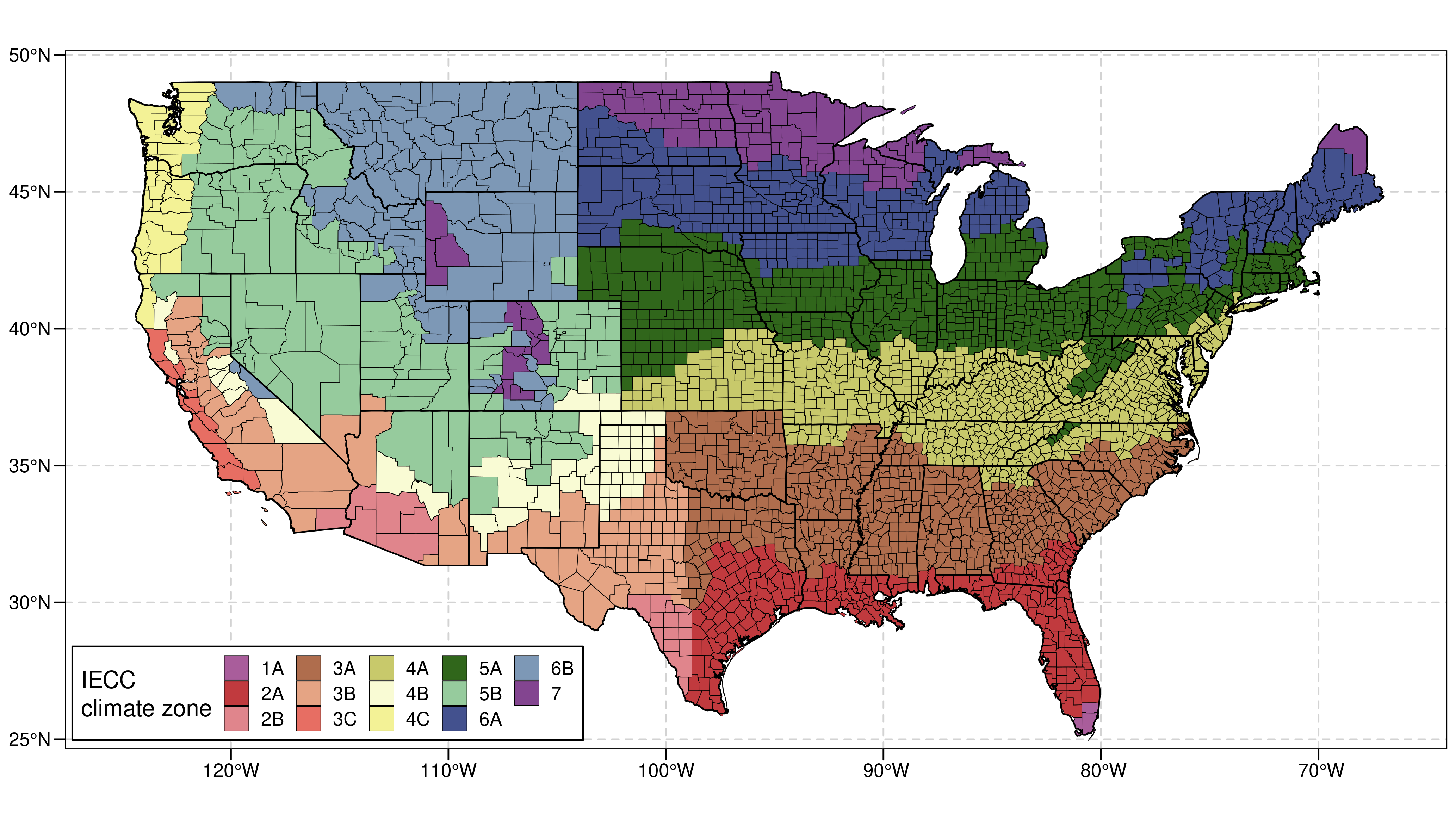}
    \caption{Geographic distribution of climate zones. State and county boundaries are outlined by solid black lines. The letters \textit{A}, \textit{B}, and \textit{C} stand for \textit{moist}, \textit{dry}, and \textit{marine} regions. The numerical numbers from 1 to 8 generally correspond to areas in need of more cooling (being hot) to more heating (being cold).}
    \label{fig:ClimateZones}
\end{figure}

It is important to evaluate and compare power production and the predictability under different climatological conditions. This study adopts the climate zones defined in the \gls{IECC} as the criterion to determine clusters within \gls{CONUS}. \gls{IECC} is a set of building codes define by the International Code Council, funded by the U.S. Department of Energy. It's main purpose is to establish the minimum design and construction requirements for energy efficiency. It assigns codes at a county level, and the code typically consists of two part, a numeric digit and a letter. The number is determined based on the thermal criteria which is the accumulated daily temperature anomaly from a predefined temperature threshold. Codes from 1 to 8 typically correspond to places from being very hot (in need of cooling) to being very cold (in need of heating). The other part of the code consists of either A, B, or C, corresponding the major climate types of moist, dry, and marine types\footnote{A rigorous definition of \gls{IECC} climate zone codes can be fount at \url{https://www.homerenergy.com/products/pro/docs/3.13/iecc_climate_zones.html}.}. These classification considers both the temperature and precipitation. In the case of \gls{CONUS} U.S., thermal conditions changes longitudinal while climate types generally change latitudinally.

\section{Methodology}
\label{sect:method}

\gls{PV} energy production is closely coupled with meteorological conditions (in particular \gls{GHI} and temperature) and solar panel configuration (panel type, location, health and installation). A complete predictability assessment of the \gls{PV} energy production essentially needs to examine both the meteorological factors from weather forecasting and the engineering factors from solar panel performance. This section introduces the framework designed for assessing the large scale predictability of \gls{PV} energy production with the multivariate \gls{AnEn} and a simulated performance system.

\subsection{Analog Ensemble and the Extension to Multivariate Forecasts}
\label{sect:MultiAnEn}

Ensemble simulation is an effective approach to assess the predictability of a weather process. However, generating a forecasting ensemble is computationally expensive. It usually involves multiple model runs, each of which can already take up significant computational resources. The \gls{AnEn} is a technique to generate forecast ensembles relying on a single run of a deterministic model and the historical observations. It is an efficient solution to ensemble generation because it saves extra simulations required by multi-initialization or multiple models. It is parallelizable and scalable because of the independent ensemble generation at an individual time and place. An in-depth description of how \gls{AnEn} works in provided below.

The \gls{AnEn} generates forecast ensembles based on the assumption that similar weather forecasts exhibit similar biases, and by using the observation, these biases can be corrected within an ensemble. The most important component of the \gls{AnEn} is a forecast similarity metric \cite{delle2013probabilistic}, defined using the following equation,

\begin{equation}
\left\|F_{t}, A_{t^{\prime}}\right\|=\sum_{i=1}^{N} \frac{\omega_{i}}{\sigma_{i}} \sqrt{\sum_{j=-\tilde{t}}^{\tilde{t}}\left(F_{i, t+j}-A_{i, t^{\prime}+j}\right)^{2}},
\label{eq:sim}
\end{equation}

where $\left\|F_{t}, A_{t^{\prime}}\right\|$ represents the distance, or the dissimilarity, between a multivariate deterministic target forecast, $F_t$, at time $t$, and an analog forecast, $A_{t^{\prime}}$, at a historical time $t^\prime$. The right side of the equation is a weighted normalized average of the differences from each predictor variable. $N$ is the number of predictor variables; $\omega_{i}$ is the weight associated with the predictor $i$; and $\sigma_{i}$ is the standard deviation, as a normalizing factor, for the predictor $i$. $F_{i, t+j}$ is the target forecast at time $t+j$ for the predictor $i$ and similarly, $A_{i, t^{\prime}+j}$ is the historical analog forecast at time $t^{\prime}+j$ for the predictor $i$. Finally, $\tilde{t}$ is the half size of the temporal window for trend comparison. This parameter is devised to consider the forecast similarity within a small time window, rather than just at a particular time point, to improve the quality of identified weather analogs. In practice, the half size of the time window is usually set to 1 hour, meaning values from the previous, the current, and the next hours will be compared during the calculation of dissimilarity at the a current time point.

For \gls{AnEn} to work properly, a set of forecasts and the corresponding observations are required to start with. The time period of the data is usually two years or longer, with at least one year as the historical period, namely the search repository, and the rest of the time as the forecast period, namely the test period. At least one year of data in the search repository is needed to account for seasonal and annual variations and to ensure good weather analogs can be found. This requirement of the length of the historical archive is much shorter than other analog-based forecasting techniques \cite{van1994searching, hamill2006probabilistic} because, with the \gls{AnEn}, weather analogs are identified within a highly restricted spatial location, independently at each grid point, and a short period of time, usually a few hours. As a result, the degree of freedom of how forecasts can differ from each other is largely constricted both in space and time, making it easier for to identify weather analogs. But the generality still holds true where better weather analogs can be found given a longer search repository.

A slightly different setting of the search and test datasets for \gls{AnEn} is called the operational \gls{AnEn}. In this setting, the size of the search repository is not fixed. Rather, it is growing by incorporating forecasts from the test dataset as soon as the target forecast becomes historical. In other words, each target forecast in the test dataset has a different length of search history that incorporates all immediate historical forecasts. This alternative setup helps in cases where only a limited length of search history is available.

After forecast analogs have been identified for a particular target forecasts, the \gls{AnEn} constructs the final ensemble by directly using the observations associated with the forecast analogs since the forecast analogs are identified from the historical archive. The historical observations compose the \gls{AnEn} forecast ensemble. This process is repeated for all geographic locations, all target forecasts in the test dataset, and all forecast lead times to generate the complete set of forecast ensembles. In fact, this process is highly parallelizable and scalable because the ensemble generation of forecasts is independent in space and time. The workload can, thus, be ``embarrassingly'' parallelized with low communication overhead \cite{cervone_short-term_2017, huparallel}.

While the spatial and temporal independence brings great advantages in computation, however, it leads to inconsistent and sometimes unrealistic predictions when ensemble members are analyzed individually \cite{sperati2017gridded, alessandrini2020schaake}. This is because the first member in the forecast ensemble at location $i$ might not necessarily be related to the first member in the forecast ensemble at a nearby location, say $i+1$. The generation of these two ensembles are independent. The member rank follows a pure statistical process of determining the distance between the target and the analog forecasts, rather than any physical linkages or processes. As a result, plotting a geographic map using the first member from ensembles at a particular day leads to patchy and unrealistic features. This problem can be well addressed by using a form of summarization of ensemble members, e.g., the mean or the median. If the ability to analyze each ensemble member is truly desired, the \gls{AnEn} members need to be shuffled with the \gls{SS} technique  \cite{clark2004schaake, voisin2010calibration, sperati2017gridded, clemente2019extension}. Ensemble spread and standard deviation can be calculated if the uncertainty of ensembles are to be evaluated. A kernel function can also be applied to the \gls{AnEn} members to construct the \gls{PDF} for probabilistic forecasts. In practice, \gls{AnEn} forecast ensembles are usually analyzed in its deterministic or the probabilistic forms, rather than on the individual member in space and time.

The \gls{AnEn} technique for predicting a univariate target has been well studied under the presumption that observations of the variable are available. This scheme does not apply when using the \gls{AnEn} to generate forecasts for initializing the \gls{PV} power simulation because the power simulation depends on multiple variables, including \gls{GHI}, temperature, wind speed, and albedo. Simplistically, one would argue that multivariate forecasts can be generated by running the \gls{AnEn} multiple times with each generation focusing on one variable at a time. However, this argument does not stand because each \gls{AnEn} generation is independent and thus, the multivariate forecasts will again suffer from having inconsistent and unrealistic predictions in space and time. In this case, the \gls{SS} does not help because it is developed for single run of \gls{AnEn}.

In this study, we propose the multivariate \gls{AnEn} and extend the application of \gls{AnEn} to multivariate forecasts with a single pass of the \gls{AnEn} technique. The multivariate \gls{AnEn} still relies on the same similarity metric, as defined in \Eq{eq:sim}, to identify analog forecasts for target forecasts. After analog forecasts have been selected, the multivariate ensemble generation shares the same set of analog forecasts. This constrains the degree of freedom when generating multivariate \gls{AnEn} forecasts in a way that the multivariate \gls{AnEn} forecasts are only possible if they have been observed as a whole in the past.

\begin{figure}
    \centering
    \includegraphics[width=0.8\textwidth]{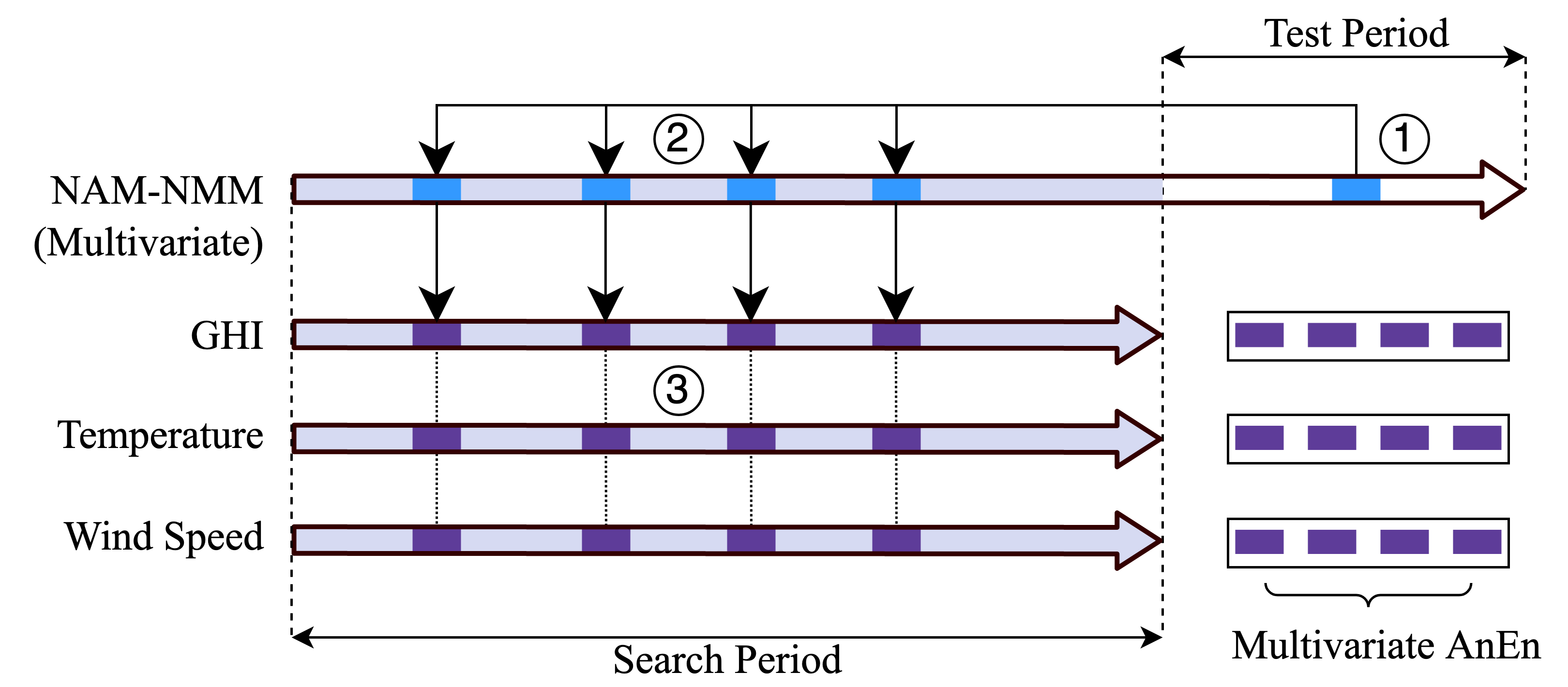}
    \caption{The schematic diagram for generating a four-member multivariate \gls{AnEn}}
    \label{fig:AnEn}
\end{figure}

\Fig{fig:AnEn} shows the schematic diagram for generating a four-member multivariate \gls{AnEn}. Forecast and observation datasets are represented as a directed arrow to indicate that their sizes might be variable, e.g., as in the operational \gls{AnEn}. The horizontal direction represents time and it is divided into the search period (as shown in grey) and the test period (as shown in white). The first arrow on from the top represents the \gls{NAM}-\gls{NMM} and by the \gls{AnEn} convention, it is a multivariate deterministic forecasts (recall the parameter $N$ in \Eq{eq:sim}). The other three arrows below represent three observation datasets that are aligned within and with the time series of \gls{NAM}-\gls{NMM}. The algorithm is described below:

\begin{enumerate}
    \item A multivariate forecast from the test period is selected for which analog forecasts are desired;
    \item Four historical forecasts from the search period with the highest similarity (the shortest distance) are identified;
    \item The associated observation values are retrieved from a particular observation dataset to form the forecast ensemble. This process is repeated for all the available observation datasets until a complete set of multivariate \gls{AnEn} has been generated.
\end{enumerate}

The parameterization of the multivariate \gls{AnEn} stays the same with the univariate \gls{AnEn}, e.g., having an 1-hour window half size and favoring a longer search period, with one exception for predictor weights. Usually, predictor weights can be optimized based on an error metric of the variable of prediction, namely the predictand, e.g., the \gls{RMSE} or the \gls{CRPS}. But the optimization objective is slightly different in the case of the multivariate \gls{AnEn} because each predictand might be associated with a different set of optimized weights if only conditioned on the individual predictand. In other words, weights for predicting \gls{GHI} could be very different from weights for predicting wind speed. We, therefore, evaluate the performance of predictor weights based on the forecast error of the final product, the simulated \gls{PV} power production, by running an power simulation system with the \gls{AnEn} forecasts.

\subsection{Scalable Ensemble Simulation for Photovoltaic Energy Production}

The most critical parameter for \gls{PV} power simulation is \gls{GHI}, which captures cloud coverage and precipitation, while the power production also depends on the configuration of solar panels and other weather variables including ambient temperature and wind speed. \Sect{sect:MultiAnEn} discusses the solution for generating multivariate weather ensembles and therefore, this section describes the later half of the power simulation workflow and running the workflow at scale.

There are a variety of software packages available for \gls{PV} system performance simulation, including \textit{PVsyst}, \textit{SAM}, \textit{PVWatts}, and \textit{PV\_LIB}. The open-source Python package \textit{PV\_LIB} is chosen by this study for several reasons. First, \textit{PV\_LIB} is an open-source package that allows in-depth scrutiny of intermediary model results and configuration and the source code is also easily available for studying purposes. It modulizes panel configurations so that it is convenient to carry out comparative analysis using varying models and assumption. Finally, its well-designed \gls{API} enables straightforward integration into external workflows, in our case, the ensemble simulation of \gls{PV} power production with \gls{AnEn}.

In general, there are five steps for simulating the \gls{PV} power production at a given location and time. The steps are described in the following paragraphs. 

\textit{Extraterrestrial Irradiance Calculation:} First, the astronomical position of the Sun relative to a particular location on the surface of the Earth needs to be calculated. This relative position can be used to calculate the extraterrestrial irradiance as well as the solar zenith angle. The process is largely deterministic due to the well established moving trajectories of planets.

\textit{\gls{GHI} Decomposition:} Second, the \gls{GHI} needs to be decomposed into the \gls{DNI} and the \gls{DHI}. The \gls{DNI} is the amount of solar radiation per unit area reaching a surface that is positioned perpendicular to the sunlight coming straightly from the sun. In production, this parameter can be further optimized by adopting the solar tracking technology where the surface of the panel is adjusted in real time to directly face the sun. The \gls{DHI} is the amount of radiation per unit area reaching a surface that does not follow a direct path but rather has been scattered by molecules or particles in the atmosphere. The three variables are interdependent and any of the variables can be derived from a combination of the other two. The DISC model is used to estimate the \gls{DNI} from the \gls{GHI}. The DISC model calculates a empirical relationship between the \gls{GHI} and the \gls{DNI} based on the global and direct clearness indices.

\textit{Incident Irradiance Estimation:} After the decomposition of \gls{GHI} from the atmosphere, these estimates need to be converted to their corresponding in-plane irradiance components. They are sometimes referred to as the plane-of-array irradiance or the incident irradiance. This calculation is necessary to determine the contributions of various components to the final power output of a \gls{PV} system.

\textit{Cell Temperature Estimation:} \gls{PV} cell temperature needs to be estimated from the weather conditions because it affects the efficiency of solar panels. Cell temperature is highly correlated with the ambient temperature but also affected by materials, e.g. glass or polymer, and wind speed.

\textit{Power Production Estimation:} Finally, the power output can be calculated based on the plane-of-array irradiance, the \gls{PV} cell condition, and the specifical \gls{PV} system configuration, like the installation and panel tilt. Since this study investigates the predictability of different solar panels, rather the optimization of a particular panel setup, it is assumed that solar panels are always facing upwards parallel to the ground. 

See \App{app:core} for a code snippet that implements the above workflow with \textit{PV\_LIB}. It provides an example to simulate power output at a given location and time\footnote{More tutorials and code examples on \textit{PV\_LIB} can be found at \url{https://pvlib-python-dacoex.readthedocs.io/en/pvsystem_tutport/usage.html}}.

There is, however, one important remark regarding integrating the aforementioned workflow into the ensemble simulation. If the complete process is carried out for each forecast ensemble separately, the calculation of extraterrestrial irradiance at a given location and time could be repeated because multiple ensembles are available. The calculation roughly takes up 8\% of the total runtime of the minimal workflow with variation depending on the platform and therefore, the computation is non-trivial. In the proposed scalable workflow, the solar position and the extraterrestrial irradiance was pre-computed only once for all locations and times of predictions, and these pre-computed results were then reused during the actual power simulation stage.

\begin{figure}
    \centering
    \includegraphics[width=0.8\textwidth]{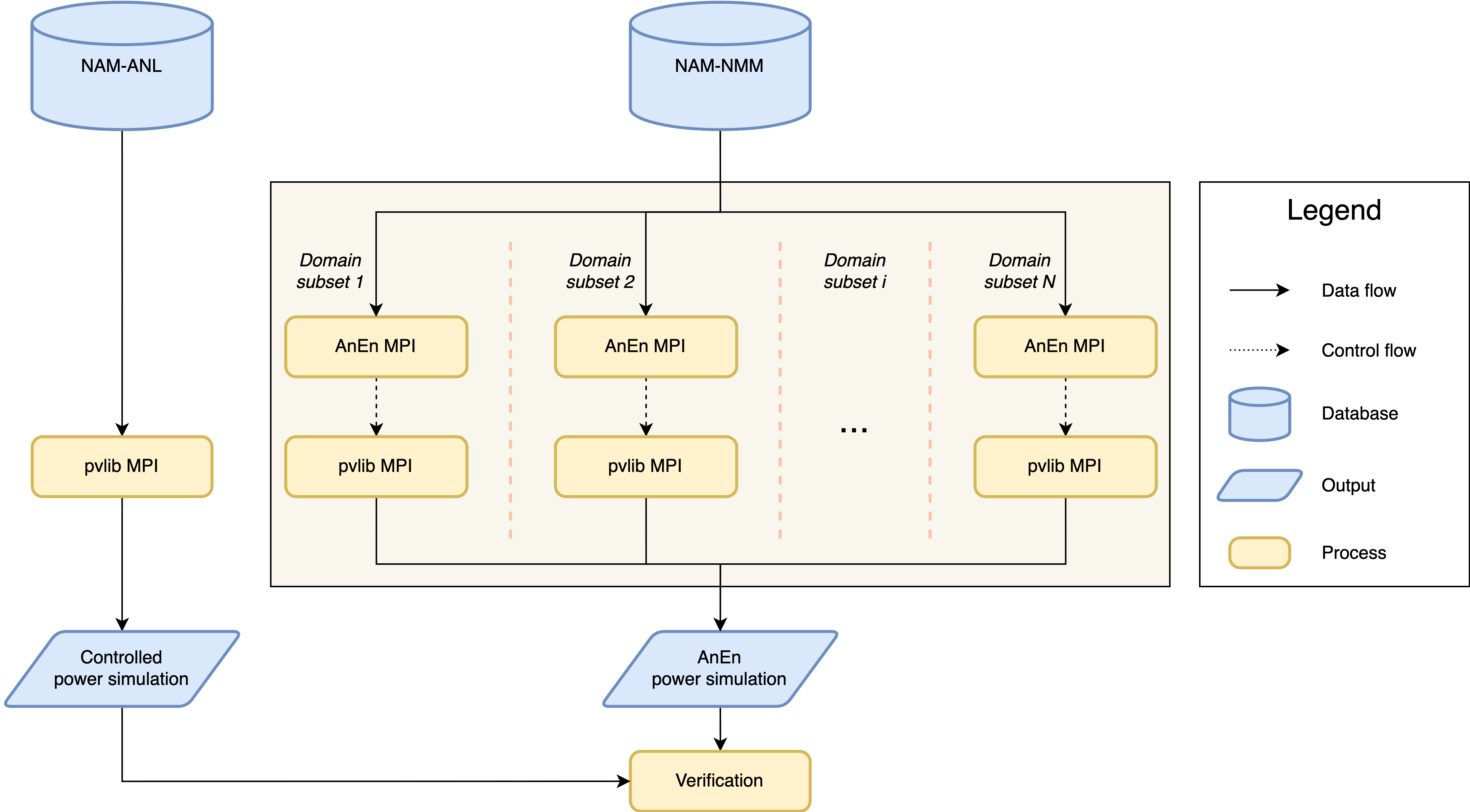}
    \caption{The schematic diagram for simulating \gls{PV} power production with \gls{AnEn}}
    \label{fig:pvsim}
\end{figure}

\Fig{fig:pvsim} shows the proposed framework for integrating the multivariate \gls{AnEn}, as implemented by \gls{PAnEn} \cite{huparallel}, and the performance simulation system, as implemented with \textit{PV\_LIB} \cite{holmgren_open_2016}. It consists of a controlled power simulation run initialized with \gls{NAM}-ANL and a batch of experimental power simulation runs initialized with \gls{AnEn} and \gls{NAM}-\gls{NMM}. This workflow is highly scalable because the generation of \gls{AnEn} can be partitioned into geographic sub-domains and run concurrently. The computation for each sub-domain can be done independently, leading to a low overhead cost of communication. This parallelization is made possible thanks to the de facto standard of \gls{MPI} for distributed computing and the workload manager, RADICAL \gls{EnTK}, developed by the RADICAL Research Group at Rutgers University. An detailed discussion on the \gls{EnTK} is provided in \Sect{sect:entk}.

\begin{sidewaystable}[htp!]
\caption{Selected solar modules for simulation. Rows are sorted based on the \gls{STC} rating decreasing}
\centerline{
\scalebox{0.9}{
\begin{tabular}{ccrcrrrr}
\hline\hline
\textit{PV\_LIB} ID & Code & Area ($m^2$) & Material & \# cells in series & Year & STC rating ($W$) & Efficiency ($\%$) \\ \hline
SunPower\_128\_Cell\_Module\_\_2009\_\_E\_\_ & SP128 & 2.144 & c-Si & 128 & 2009 & 400 & 18.7 \\
SunPower\_SP305\_GEN\_C\_Module\_\_2008\_\_E\_\_ & SP305 & 1.63 & c-Si & 96 & 2008 & 305 & 18.7 \\
Silevo\_Triex\_U300\_Black\_\_2014\_ & STU300 & 1.68 & c-Si & 96 & 2014 & 300 & 17.8 \\
Suniva\_Titan\_240\_\_2009\_\_E\_\_ & ST240 & 1.643 & c-Si & 60 & 2009 & 240 & 14.7 \\
BP\_Solar\_BP3232G\_\_\_2010\_ & BP3232G & 1.6068 & c-Si & 120 & 2010 & 230 & 14.8 \\
Sharp\_ND\_216U1F\_\_2008\_\_E\_\_ & ND216U1F & 2.63 & mc-Si & 60 & 2008 & 215 & 8.3 \\
Solar\_Frontier\_SF\_160S\_\_2013\_ & SF160S & 1.22 & CIS & 172 & 2013 & 160 & 13.2 \\
Kyocera\_Solar\_KD135GX\_LP\_\_2008\_\_E\_\_ & KD135GX & 1.002 & mc-Si & 36 & 2008 & 135 & 13.47 \\
Kyocera\_Solar\_KC85T\_\_2008\_\_E\_\_ & KC85T & .656 & mc-Si & 36 & 2008 & 85 & 13.4 \\
First\_Solar\_FS\_272\_\_\_2009\_ & FS272 & .72 & CdTe & 116 & 2009 & 72.5 & 10.07 \\
Kyocera\_Solar\_KS20\_\_2008\_\_E\_\_ & KS20 & .072 & mc-Si & 36 & 2008 & 20 & 28 \\
\hline\hline
\end{tabular}}}
\label{tab:Panel}
\end{sidewaystable}

Finally, the workflow is repeated for 11 different solar panel types. There are, in total, 523 built-in solar modules in \textit{PV\_LIB} (as of \textit{pvlib} v0.8.1) provided by the Sandia Module database and 21,535 from the Clean Energy Council module database. Even with access to supercomputers and a large allocation, it is prohibitive to simulate all the modules.  It is also unnecessary, because while all panels are different, they can be clustered into groups of similarly behaving panels.

Clusters were generated using the size of the solar panel, the total number of \gls{PV} cells in a series, and the \gls{STC} power rating for the panels in the Sandia Module database, and we carried out a clustering analysis to identify 11 module clusters. We then select the module with the maximum \gls{STC} rating from each cluster. A summary of the selected solar modules is provided in \Table{tab:Panel}.

\section{Results and Discussions}
\label{sect:results}

Hourly day-ahead \gls{PV} power production simulation for 2019 with a spatial resolution of 12 km has been carried out over \gls{CONUS}. Because the power production at night is zero, only daylight hours were used. This leads to a different number of daily forecasts because for each location the length of the days varies throughout the year.  

\gls{AnEn} has 21 members and uses a two-year search period start at January 1, 2017, with the operational mode. Please recall that, in the operational mode, the search period is accumulated as the test time moves forward. Each of the 21 members is used to simulate power production. In the next section, results are shown for the weight optimization of the multivariate \gls{AnEn} generation and the power simulation ensembles.

\subsection{Weight Optimization for Multivariate Analog Ensemble Generation}
\label{sect:weight}

One of the important input for \gls{AnEn} is the set of predictor weights. Previous research employed a brute-force search that experiment different combinations of weights for each predictor, say from 0 to 1 with an increment of 0.1. This is, by nature, a computationally expensive task. Because all weights are normalized, a simple optimization is to force all weights to add up to 1. 

In this study, the weight optimization follows this practice. However, having one set of weights (hereafter \gls{EW}) for the entire region of \gls{CONUS} is not optimal because of the large domain coverage which calls for different optimal weights. On the other hand,  optimizing for each of the 56,776 grid points is too computationally taxing. As a compromise, two alternative approaches are discussed to set weights for each of the grids. 

\begin{enumerate}
    \item The first approach optimizes weights from a sample of equally-spaced grids within \gls{CONUS} and each grid is assigned the weights of the nearest sample point (hereafter \gls{NN}). The sample points have a latitudinal spacing of 4.5° and a longitudinal spacing of 3.5°. 
    
    \item The second approach uses a hierarchical clustering algorithm based on four features, including orography, annual \gls{GHI}, temperature, and the average cloud cover.  The rationale is that these four parameters inform the geographic and meteorological regimes (hereafter \gls{RB}). 
\end{enumerate}

For both methods \gls{CRPS} is the metric or objective function being minimized, and it is computed at the center points of the \gls{NN} sample grids, or for \gls{RB} at a random sample of points from each cluster using data for 2018.  The \gls{CRPS} is calculated based on the actual power generation from the \gls{PV} system rather than on the \gls{GHI}, which is the real quantity of importance for this research.  Rather than trying all 11 panels, the optimization is performed only on a single  \gls{PV} module, specifically the Silevo Triex U300.  This was chosen due to its relatively large power capacity of 300 $W$ and its popularity in real world installations.

\begin{figure}
    \centering
    \includegraphics[width=0.9\textwidth]{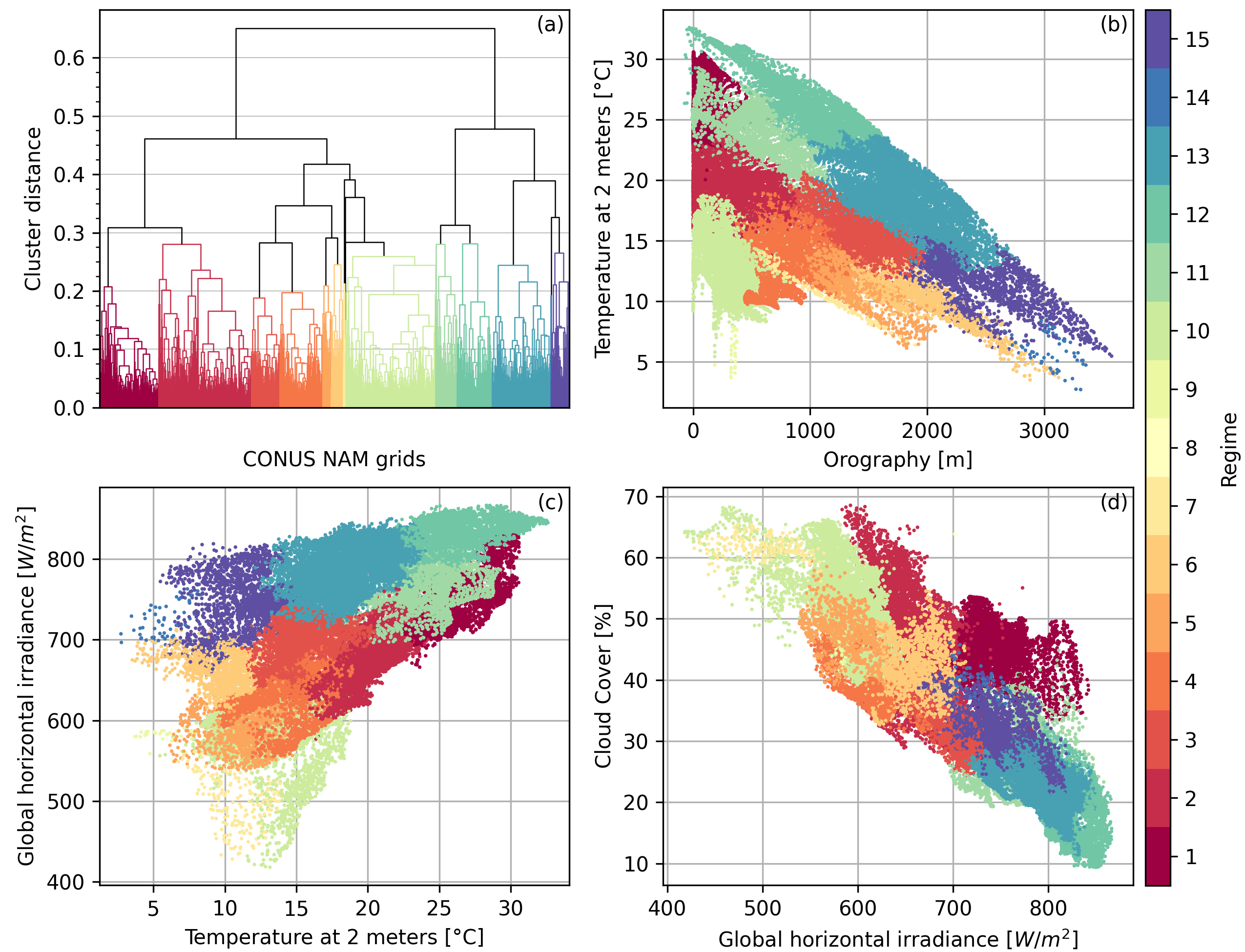}
    \caption{Hierarchical clustering of \gls{NAM}-\gls{NMM} grid points: a dendrogram representing the calculated hierarchical clusters (a) and scatter plots of predictors used to build the hierarchical clusters (b, c, d). Individual clusters are color coded.}
    \label{fig:hclust}
\end{figure}

\Fig{fig:hclust} shows how the clusters are defined. The hierarchical clustering algorithm is alternative to the well known K-Mean algorithm, and it differs in not requiring the number of clusters as input. Rather, the number of clusters is based on point distances.

There are two sub-categories of the hierarchical clustering, divisive and agglomerative \cite{pedregosa2011scikit, nielsen2016introduction}. In agglomerative clustering, the algorithm starts by assigning each grid point to its own cluster, and then follows an iterative process where the two closest clusters are merged. Different metrics can be used to define the distance between clusters, e.g., using the shortest pair of points from two clusters or using the average distance between two clusters. The first definition is used when there are clear-cut shapes in the data; whereas the second definition is used when geographic locations are naturally correlated and variables tend to change smoothly \cite{ferstl2016time}. Because of the high spatial correlation between grid points, this research uses the average distance between two clusters.

An effective way to visualize the process of the hierarchical clustering is through a dendrogram (\Fig{fig:hclust}a). Clusters that are closer to each other are drawn closer along the horizontal axis. The clustering is an iterative process, meaning that, by the end of the process, there will only be one big cluster consisting of all the grid points. The height of a horizontal line connecting two clusters indicates the distance between merger clusters. This figure helps to decide the number of final clusters by providing information on how close the clusters are and how many points each cluster would have. We therefore, generated 15 clusters. \Figs{fig:hclust}a, b, c show the relationship between variables used for clustering. The general correlation between variables is well-expected, e.g., a negative correlation between temperature and orography, or cloud cover and \gls{GHI}; and a positive correlation between \gls{GHI} and temperature. The clusters separate the regimes effectively, for example \#15 corresponds to a regime of high elevation, low temperature, and high level of \gls{GHI}.

\begin{figure}
    \centering
    \includegraphics[width=\textwidth]{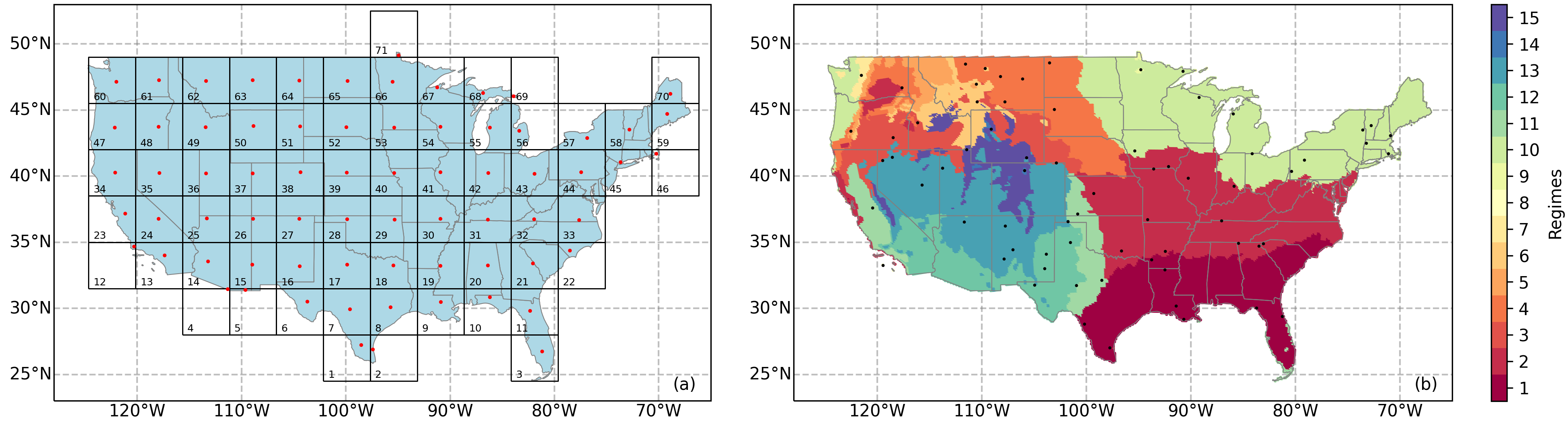}
    \caption{Geographic distribution of clusters determined from \gls{NN} (a) and \gls{RB} (b). There are 71 clusters and 15 clusters in \gls{NN} and \gls{RB} respectively. Samples points are shown in red (a) for \gls{NN} and in black (b) for \gls{RB}. \gls{NN} clusters are labeled to the bottom left of each cluster. \gls{RB} clusters are color coded. State boundaries are shown in grey lines.}
    \label{fig:clust}
\end{figure}

\Fig{fig:clust} arranges the \gls{NN} and the \gls{RB} clustering within the geographic context. \Fig{fig:clust}a shows the equally-spaced grid for each cluster. The red point shows the centroid location of the overlapped region for which weights will be optimized. \Fig{fig:clust}b shows the \gls{RB} clusters and the black points show the sample locations where the weights are optimized. The total number of sample points from two approaches remains similar, with 71 for the \gls{NN} and 70 for the \gls{RB}. \Fig{fig:clust}b shows consistency with the \gls{IECC} climate zones, shown in \Fig{fig:ClimateZones}. However, the determination of regimes is more dependent on solar-related features, while the climate zones are mostly defined by temperature and precipitation. 

\begin{figure}
    \centering
    \includegraphics[width=\textwidth]{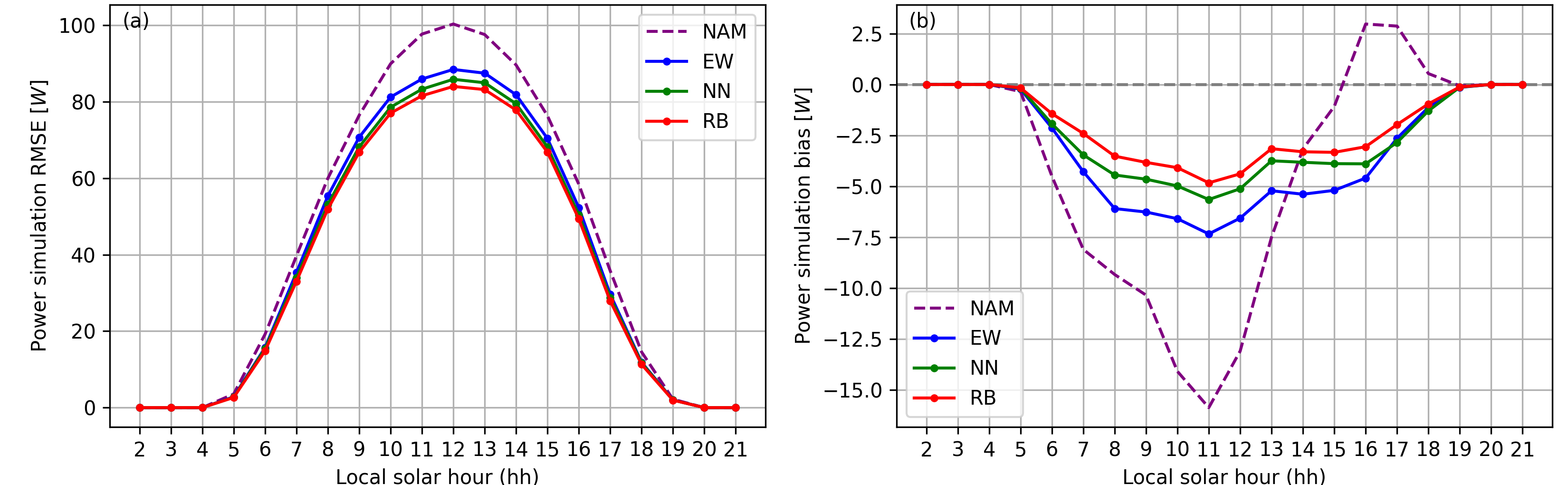}
    \caption{\gls{RMSE} (a) and bias (b) of \gls{PV} power simulation by forecast lead times for 2019 from four methods of comparisons: \gls{NAM}, \gls{EW}, \gls{NN}, and \gls{RB}. Forecast lead times from different geographic locations have been aligned with local solar time so that 12 is always the local solar noon.}
    \label{fig:veriTime}
\end{figure}

A verification along space and time was used to evaluate the effectiveness of the optimized weights. \Fig{fig:veriTime} shows the \gls{RMSE} for four prediction methods, the baseline model \gls{NAM} and three different \gls{AnEn} results, each characterized by a different weighting strategies. The \gls{RMSE} is calculated based on power simulation, meaning that both \gls{NAM} and \gls{AnEn} forecasts were used as input to the panel simulator. Model analysis has also been fed into the simulator to generate the most realistic version of what actually happened as the ``ground-truth''.

\gls{NAM} forecasts are computed in UTC times, which is not convenient when comparing large regions. Because \gls{CONUS} effectively spans over four UTC time zones, significant artifacts are introduced when comparing far away regions like the East and West coasts. For example, generating forecasts for 17h00 UTC effectively means comparing forecasts for noon local time in New York with 09h00 local time in Los Angeles.

To avoid these inconsistencies, all computations are made in local time.  However, rather than using the UTC time zone maps to carry out the time conversion, which adds an artificial transformation between solar time and a geographical region, the real solar noon is computed at each location, and the forecast lead times are realigned so that the forecast noon is the closest solar noon.  Solar noon is the time when the Sun passes a location's precise meridian and reaches its highest elevation in the sky. This time corresponds to the highest level of solar extraterrestrial irradiance, and  on a clear sky day also corresponds to the highest level of surface irradiance. Because \gls{NAM} provides hourly forecasts, the time difference between the solar noon and the forecast noon is at most $\pm$30 minutes.  

\Fig{fig:veriTime}a shows the \gls{RMSE} averaged across \gls{CONUS} as a function of lead time for the \gls{AnEn} ensemble means of \gls{EW}, \gls{NN}, and \gls{RB}, and for the deterministic forecasts of \gls{NAM}-\gls{NMM}. On average, the \gls{RMSE} of \gls{NAM}-\gls{NMM} peaks at solar noon with a value of 100 $W/m^2$. \gls{EW} \gls{AnEn} reduces the prediction error by 11.83\%, from 100 to 88.42 $W/m^2$.  However, using only equal weights for searching weather analogs is sub-optimal. \gls{AnEn} \gls{NN} and \gls{RB} show results from the two optimizing schemes tested.  \gls{NN} reduces 14.41\% of the prediction error of \gls{NAM}-\gls{NMM} from 100 to 85.85 $W/m^2$ and \gls{RB} reduces it by 16.26\% from 100 to 83.99 $W/m^2$. While the improvement is largest at solar noon, it follows the same trend for the other lead times, thus with \gls{RB}, \gls{NN}, \gls{EW} and \gls{NAM} consistently ordered from lowest to highest error (\Fig{fig:veriTime}a)

\Fig{fig:veriTime}b shows the prediction bias of the four methods. The bias is a verification metric that captures whether a weather model systematically over- or under-predicts. Ideally predictions should have a bias of zero, meaning that the forecasted ensemble mean always overlaps with the real mean. A positive bias, also referred to as a high bias, means that the forecasted ensemble mean is higher than the real mean.  A negative bias, also referred to as low bias, is the opposite.

In terms of solar irradiance forecasts, \gls{NAM}-\gls{NMM} generally exhibits a negative bias during most of the day time but shifts to a positive value during afternoons. The peak value is a negative bias that occurs one hour before solar noon. Because the \gls{RMSE} is increasing during this time while the bias improves slightly, an increase in random error is offsetting the bias improvement during this period of time.

All three modes of \gls{AnEn} have better biases. At solar 11:00, when the bias of \gls{NAM}-\gls{NMM} is worst, the four methods have the bias of -15.87 $W/m^2$ (\gls{NAM}), -7.34 $W/m^2$ (\gls{EW}), -5.65 $W/m^2$ (\gls{NN}), and -4.83 $W/m^2$ (\gls{RB}).

\begin{figure}
    \centering
    \includegraphics[width=\textwidth]{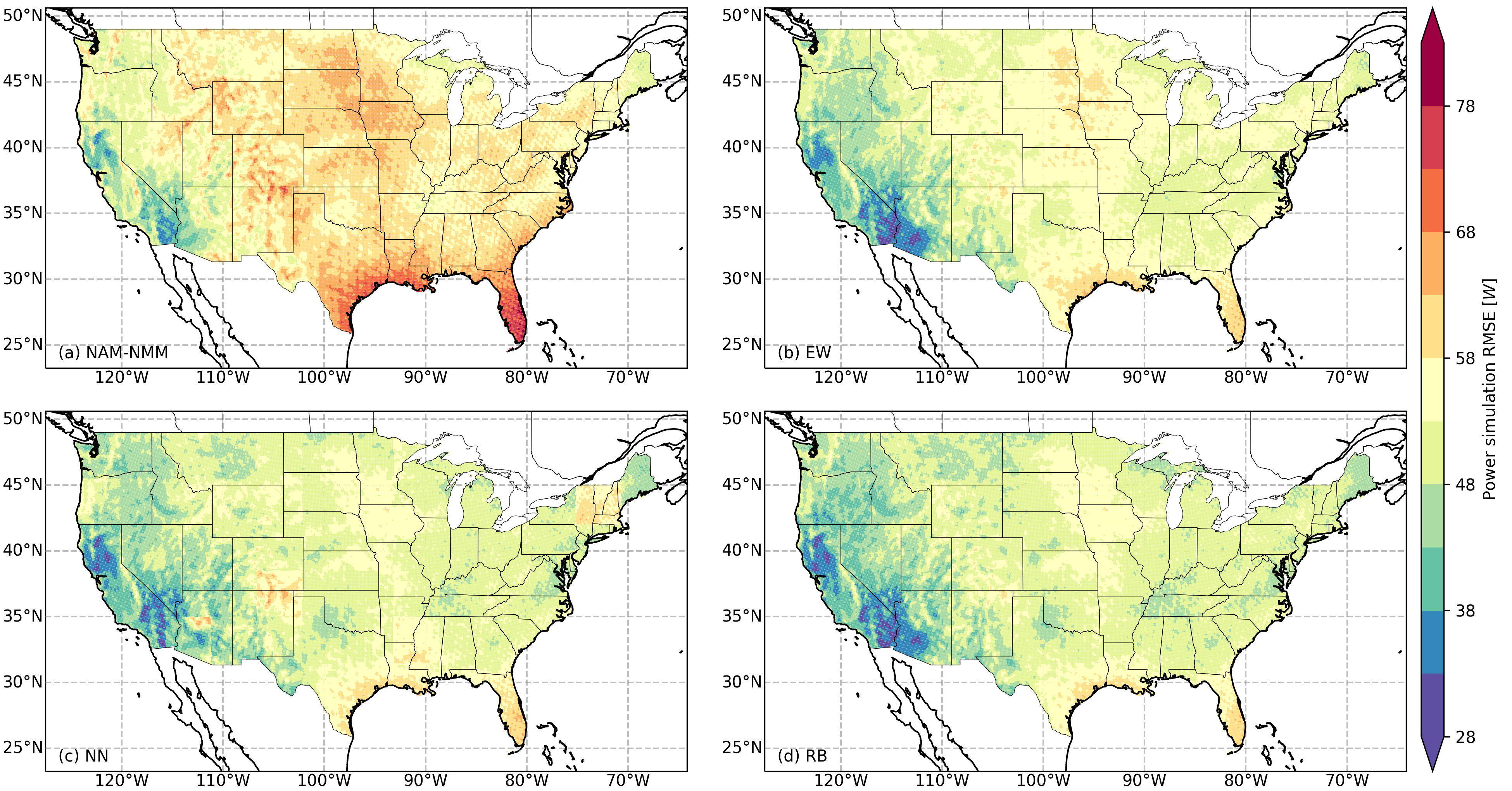}
    \caption{\gls{RMSE} of \gls{PV} power simulation averaged from all lead times for 2019 from four methods of comparisons: \gls{NAM} (a), \gls{EW} (b), \gls{NN} (c), and \gls{RB}. Coasts and state boundaries are shown in solid lines.}
    \label{fig:VeriLocations}
\end{figure}

\Fig{fig:VeriLocations} shows the \gls{RMSE} maps of the entire domain for the four methods. Each pixel shows the yearly average of \gls{RMSE} for all daylight lead times. Prominent in \Fig{fig:VeriLocations}a are two high error regions over Florida and Southern Texas. \Figs{fig:Analysis2018}c, d show that these two regions generally receive a high level of solar irradiance and total cloud cover. High levels of power production can be achieved because of the high solar irradiance, but the prediction error is also high due to high cloud cover. 

However, another region is the Pacific Northwest which also has high cloud cover, instead, shows visibly lower prediction errors than that for Texas and Florida. This is mostly likely due to the different amount and variability of cloud cover. Clouds over the Pacific Northwest are regular and persistent, which is in part reflected by lower solar irradiance values (\Fig{fig:Analysis2018}c). The absolute error is therefore smaller because the magnitude of the power output is never as high as in Florida and Texas. Another feature to note in \Fig{fig:VeriLocations}a is the high accuracy over Arizona and the Southern California. These regions have high solar irradiance and low cloud cover, making them easier to predict. 

\Figs{fig:VeriLocations}b, c, d show respectively the \gls{RMSE} for \gls{EW}, \gls{NN}, and \gls{RB}. All three methods show lower errors compared to \gls{NAM}-\gls{NMM} overall, as well as in specific areas. This can be attributed to the independent search of weather analogs with a fine spatial scale at each grid point. It allows the weather analogs to adapt to local weather regimes and correct the forecasts accordingly.

\Fig{fig:VeriLocations} reveals one systematic problem with the \gls{NN} approach. Recall from \Fig{fig:clust}a that, due to the computational limitations, it is unfeasible to run weight optimizations for each grid point. Only the center of grid points are used to compute the optimal weights, which are then reused for the entire grid of roughly 4.5°x3.5°. If a non-representative sample point for the entire grid is selected, incorrect weights are used for several points. 

Its negative impact on prediction error is most noticeable for the 14th, 19th, 27th, and 58th regions in \Fig{fig:clust}. These regions have larger \gls{RMSE}, surrounded by areas with noticeably lower errors. On the other hand, \gls{RB}, that uses four variables (topography, solar irradiance, temperature, and cloud cover) to define various regimes, has a higher homogeneity in term of power production, which results in a smoother \gls{RMSE} surface (\Fig{fig:VeriLocations}d). 

\begin{figure}
    \centering
    \includegraphics[width=\textwidth]{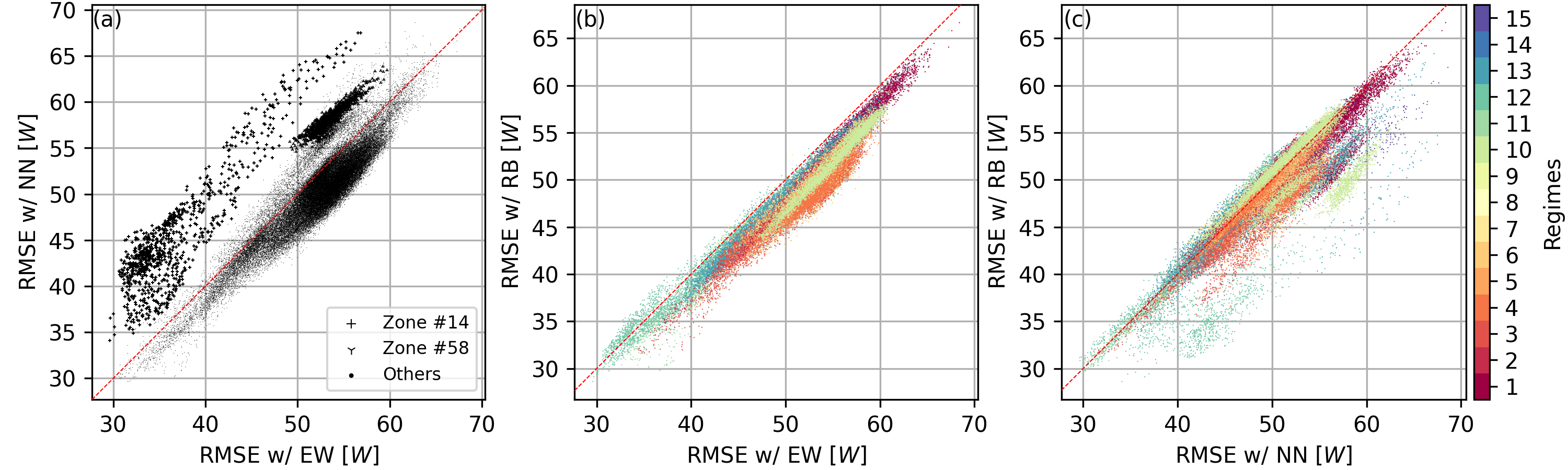}
    \caption{Scatter plots showing \gls{RMSE} of \gls{PV} power simulation for 2019: \gls{NN} v.s. \gls{EW} (a), \gls{RB} v.s. \gls{EW}, and \gls{RB} v.s. \gls{NN}. Each point in the scatter plots represents the verification statistic of a geographic location (a \gls{NAM} grid point) calculated from all lead times during 2019. Two representative zones are highlighted with different point types in (a) with the index number corresponding to \Fig{fig:clust}a. Points are color coded based on regimes in (b, c) with a color mapping in consistent with \Fig{fig:clust}b.}
    \label{fig:scatter}
\end{figure}

\Fig{fig:scatter} shows scatter plots with pairwise comparisons for all modes of \gls{AnEn}, along with a diagonal reference line.  Points below the diagonal indicate that the method on the vertical axis is better. \Fig{fig:scatter}a compares \gls{EW}, on the horizontal axis, with \gls{NN}, on the vertical axis. The majority of the points lie below the diagonal line, suggesting that \gls{NN} outperforms \gls{EW}. However, points from the 14th and the 58th (\Fig{fig:veriTime}a) grids are highlighted using different shapes, because they all show an inverse trend, indicating that the weight optimization failed for these regions. 

These issues are likely to be solved by increasing the number of  clusters, and reducing their geographical size. However, this will significantly increase computation time, which is what the method is trying to avoid.  Without increasing the sample points, \Fig{fig:scatter}b shows the comparison between the \gls{EW} on the horizontal axis and the \gls{RB} on the vertical axis. Please recall that the number of samples points are similar with 71 for the \gls{NN} and 70 for the \gls{RB}. The majority of points lie below the diagonal indicating the clear outperformance of \gls{RB} compared to \gls{EW}.

\Fig{fig:scatter}c compares the \gls{NN} and \gls{RB}. The difference between these two methods are smaller than the previous comparisons because both methods (\gls{NN} and \gls{RB}) involve grid search but with differences in how regions are defined and samples are drawn. Yet, \gls{RB} shows its outperformance over the \gls{NN} for the regimes 10, 11, and 12 which overlaps with the \gls{NN} clusters where non-representative sample points are chosen.

\gls{RB} also demonstrates patterns of errors related to the spatial characteristics of the regimes. For example, the regime 1 covers the Southeastern \gls{CONUS} associated with high solar irradiance and and middle-to-high cloud cover. Points are clustered, in \Fig{fig:scatter}b, at the right tail of the spectrum, having a high error of power simulation. On the other side, regime 13 and 12 cluster at the left tail of the spectrum, having a low error of power simulation. These regimes are typically associated with middle-to-high solar irradiance but very low cloud cover. The clustering of error within \gls{RB} regimes indicates the effectiveness of the regimes to separate different error patterns associated with \gls{NAM}-\gls{NMM}, and as a result, the predictor weights can be optimized focusing on correcting a particular type of error.

\begin{table}
\caption{Summary statistics calculated based on the hierarchical regimes define by the \gls{RB}. Power simulation \gls{RMSE} is calculated at local solar noon averaged across 2019 and from all grid points. \text{*} indicates that the \gls{RMSE} of the \gls{RB} is significantly different from that of \gls{NAM}-\gls{NMM}, the \gls{EW}, and the \gls{NN} at the level of $p=0.05$.}
\centerline{
\scalebox{0.85}{
\begin{tabular}{crrrrrrr}
\hline\hline
\multirow{2}{*}{\begin{tabular}[c]{@{}c@{}}RB\\ Regime\end{tabular}} & \multicolumn{4}{c}{RMSE at Solar Noon ($W$)} & \multirow{2}{*}{\begin{tabular}[c]{@{}c@{}}GHI at\\ Solar Noon ($W/m^2$)\end{tabular}} & \multirow{2}{*}{\# Grids} & \multirow{2}{*}{\# Samples} \\ \cline{2-5}
 & NAM & EW & NN & RB &  &  &  \\ \hline
12 & 86.641 & 73.424 & 74.861 & \textbf{71.871} & 362.303 & 4225 & 5 \\
3 & 91.641 & 82.355 & 80.444 & \textbf{76.961} & 301.824 & 3411 & 4 \\
11 & 94.087 & 82.926 & 78.492 & \textbf{77.314} & 319.546 & 2615 & 3 \\
13 & 92.382 & 82.056 & 81.237 & \textbf{79.817} & 342.535 & 7105 & 8 \\
7 & 93.546 & 84.277 & \textbf{79.760} & 82.431 & 176.769 & 208 & 1 \\
5 & 94.271 & 87.009 & 83.442 & \textbf{82.604} & 234.463 & 1045 & 2 \\
4 & 100.165 & 89.482 & 85.378 & \textbf{83.068} & 273.874 & 5188 & 6 \\
6 & 98.555 & 89.330 & 86.413 & \textbf{84.098} & 267.284 & 1477 & 2 \\
2 & 101.926 & 90.213 & 86.498 & \textbf{85.399} & 272.640 & 11221 & 13 \\
15 & 102.211 & 91.227 & 88.748 & \textbf{87.301} & 318.927 & 2160 & 3 \\
10 & 101.705 & 93.053 & 89.207 & \textbf{87.607} & 239.381 & 10884 & 12 \\
1 & 112.170 & 92.922 & 91.834 & \textbf{89.009} & 289.397 & 7134 & 8 \\
9 & 110.560 & 96.758 & \textbf{90.027} & 92.894 & 258.179 & 33 & 1 \\
14 & 110.357 & 95.756 & \textbf{94.004} & 94.017 & 290.165 & 68 & 1 \\
8 & 143.911 & 117.995 & 118.537 & \textbf{116.924} & 227.277 & 2 & 1 \\ \hline
Summary & 102.275 & 89.919 & 87.259 & \textbf{\text{*}86.088} & 278.304 & 56776 & 70 \\ \hline\hline
\end{tabular}}}
\label{tab:rb}
\end{table}

\begin{sidewaystable}
\caption{Summary statistics aggregated from \gls{IECC} climate zones. Power simulation \gls{RMSE} is calculated from different parts of the day and averaged across 2019 and all grid points:  morning hours from 8 AM to 10 AM, noon hours from 11 AM to 1 PM, and afternoon hours from 2 PM to 4 PM. \text{*} indicates that the \gls{RMSE} of the \gls{RB} is significantly different from its counterpart of \gls{NAM}-\gls{NMM}, the \gls{EW}, and the \gls{NN} at the level of $p=0.05$.}
\centerline{
\scalebox{0.9}{
\begin{tabular}{crrrrrrrrrrrrr}
\hline\hline
\multirow{2}{*}{\begin{tabular}[c]{@{}c@{}}IECC\\ Climate Zone\end{tabular}} & \multicolumn{4}{c}{RMSE during Morning ($W$)} & \multicolumn{4}{c}{RMSE during Noon ($W$)} & \multicolumn{4}{c}{RMSE during Afternoon ($W$)} & \multirow{2}{*}{\# Grids} \\ \cline{2-13}
 & NAM & EW & NN & RB & NAM & EW & NN & RB & NAM & EW & NN & RB &  \\ \hline
1A & 92.159 & 71.736 & 71.549 & \textbf{69.321} & 129.349 & 105.522 & 105.502 & \textbf{104.020} & 106.667 & 88.859 & 88.370 & \textbf{88.135} & 72 \\
2A & 90.326 & 77.156 & 76.485 & \textbf{74.920} & 115.530 & 96.823 & 95.377 & \textbf{93.812} & 87.083 & 76.633 & 75.332 & \textbf{74.272} & 3021 \\
2B & 72.862 & 60.536 & 66.694 & \textbf{60.457} & 88.260 & 71.653 & 77.103 & \textbf{70.901} & 65.935 & 55.248 & 60.235 & \textbf{54.357} & 1061 \\
3A & 78.575 & 71.876 & 70.402 & \textbf{68.229} & 100.953 & 88.554 & 86.707 & \textbf{83.941} & 79.173 & 71.294 & 69.726 & \textbf{67.754} & 6621 \\
3B & 71.429 & 62.419 & 60.788 & \textbf{60.062} & 88.596 & 76.456 & 74.795 & \textbf{73.735} & 67.588 & 60.503 & 58.947 & \textbf{57.821} & 4350 \\
3C & 68.007 & 58.530 & 56.144 & \textbf{55.821} & 80.987 & 73.096 & 70.623 & \textbf{70.044} & 62.498 & 54.722 & 52.545 & \textbf{51.704} & 395 \\
4A & 76.997 & 70.917 & 68.056 & \textbf{67.503} & 98.880 & 88.183 & 84.665 & \textbf{83.641} & 76.653 & 69.828 & 67.127 & \textbf{66.426} & 6049 \\
4B & 72.845 & 65.486 & 65.151 & \textbf{62.753} & 94.035 & 80.596 & 81.420 & \textbf{77.660} & 76.439 & 69.087 & 69.432 & \textbf{66.643} & 2150 \\
4C & 71.933 & 66.813 & 66.272 & \textbf{63.700} & 90.069 & 83.501 & 82.578 & \textbf{79.425} & 66.980 & 61.967 & 61.641 & \textbf{58.162} & 1139 \\
5A & 79.830 & 74.841 & 70.627 & \textbf{70.118} & 100.940 & 91.193 & 86.295 & \textbf{85.450} & 75.443 & 70.100 & 66.491 & \textbf{65.752} & 7068 \\
5B & 71.148 & 62.947 & 61.103 & \textbf{59.838} & 90.951 & 80.541 & 79.091 & \textbf{77.493} & 71.695 & 65.183 & 63.423 & \textbf{62.053} & 8931 \\
6A & 77.192 & 73.617 & 70.074 & \textbf{68.900} & 101.898 & 92.408 & 89.019 & \textbf{86.991} & 74.903 & 69.967 & 67.080 & \textbf{65.869} & 5870 \\
6B & 72.740 & 68.634 & 64.640 & \textbf{62.710} & 97.529 & 88.247 & 84.517 & \textbf{82.523} & 76.649 & 71.578 & 68.109 & \textbf{66.624} & 6518 \\
7 & 71.947 & 70.265 & 65.768 & \textbf{65.368} & 100.260 & 90.829 & 86.651 & \textbf{85.569} & 75.322 & 70.253 & 67.159 & \textbf{66.396} & 3069 \\ \hline
Summary & 76.285 & 68.269 & 66.697 & \textbf{\text{*}64.978} & 98.445 & 86.257 & 84.596 & \textbf{\text{*}82.515} & 75.930 & 68.230 & 66.830 & \textbf{\text{*}65.141} & 56314 \\ \hline\hline
\end{tabular}}}
\label{tab:iecc}
\end{sidewaystable}

Finally, \Table{tab:rb} and \Table{tab:iecc} compare the forecast error based on two different geographic clustering criteria. \Table{tab:rb} averages errors from the regimes defined by the \gls{RB}. Rows in the table are sorted based on the \gls{RMSE} of the \gls{RB} from the lowest to the highest. Verification results are shown for the solar noon only because that is genearlly when the solar irradiance and the power production reach the climax.

The \gls{RB} mostly achieves lower errors compared to the \gls{NN} except for the regimes 7, 9, and 14. \gls{RB} only has one sample point for each of these regimes, making it subject to the same issue \gls{NN} has. However, these regimes are tend to be small clusters and therefore have weaker impact to the overall predictability, unlike in the case of \gls{NN}, each sample point will have an equal impact due to the same size of each cluster. On average, at the solar noon, the \gls{RB} has the lowest prediction error. The average statistics is consistent with \Fig{fig:veriTime}a at solar noon time.

It is possible that \gls{RB} outperforms the other methods simply because the verification is aggregated in favor of how \gls{RB} is optimized. To rule out the potential bias, an independent clustering, from the \gls{IECC} climate zones, is used to average errors, as shown in \Table{tab:iecc}. The total number of grids within the \gls{IECC} climate zones is smaller (462 points, 0.8\%) than the total number of grid points within \gls{CONUS} as defined by \gls{NAM}-\gls{NMM}, because of a slight difference in the coordinates mostly along the coasts (not shown).

Similar results can be observed that \gls{RB} predominately remains the best method compared to \gls{NAM}-\gls{NMM}, the \gls{EW}, and the \gls{NN}. In general, error reduction during the noon is larger than the reduction during the morning and the afternoon. The climate zone, $3C$, is shown to have the lowest prediction error throughout the day and methods. It is, again, associated with high solar irradiance and low cloud cover.

It is therefore concluded that \gls{RB} is a more reliable and effective weight optimization method for problems where a tradeoff must be found between computing weights for a large geographic region, and limited computational power available.

\subsection{Predictability Assessment of Solar Panel Configuration}
\label{sect:assess}

\Sect{sect:weight} evaluates two approaches to weight optimization for \gls{AnEn} predictions. Results have been shown that \gls{RB} is significantly better and more reliable for a large geographic domain. In this section, predictor weights optimized by the \gls{RB} method is used to generate \gls{AnEn} forecasts. Weather ensembles are then used with 11 \gls{PV} modules (\Table{tab:Panel}) to simulate power production.

\gls{PV} systems are simulated with a 10 $KW$ capacity by the linear scaling of a particular type of panel in a series. For example, according to \Table{tab:Panel}, the panel, \textit{SP128}, has 400 $W$ capacity. Therefore, the simulated system contains 25 such panels in series to achieve the desired power of 10 $KW$. It is possible to connect \gls{PV} modules in parallel to account for shading, but it requires extra wiring. Currently, connecting modules in series remains the most common practice because it generates maximum power output while at the same time being cheaper to install. In this \gls{PV} system, we do not compensate for scaling and conversion losses, which depend on the type of inverter used. 

Each \gls{PV} system is run with hourly forecasts from \gls{AnEn} for 2019. Since \gls{AnEn} forecasts have 21 members, power simulation also has 21 members for each of the 11 systems.

\begin{figure}
    \centering
    \includegraphics[width=\textwidth]{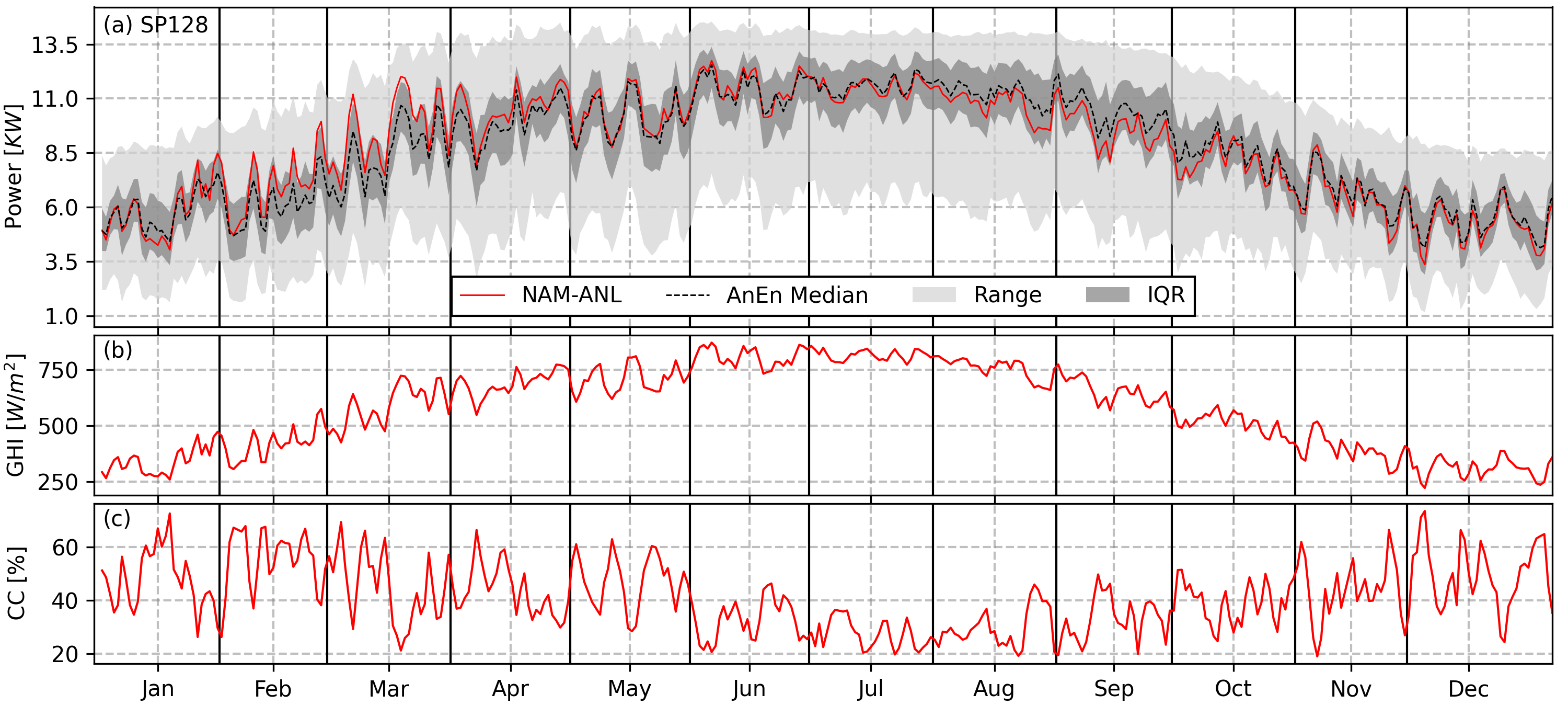}
    \caption{21-member power simulation ensemble from the 10 $KW$ \gls{PV} system with the module, $SP128$ for 2019. Only the power production at noon for each day is shown. \gls{NAM}-\gls{NMM} is shown in solid red; the ensemble median is shown in dashed black; the shaded colors indicate the ensemble range and the Inter-Quantile Range (IQR). (b, c) show the corresponding solar irradiance and cloud cover from \gls{NAM}-\gls{NMM}.}
    \label{fig:SP128}
\end{figure}

\Fig{fig:SP128} gives an example of the simulated power ensemble from \textit{SP128}. The figure shows the power production at solar noon on each day in 2019. Values are averaged across all \gls{CONUS} grid points. \Figs{fig:SP128}a, b show a high positive correlation. However, the \gls{GHI} reaches its climax during July while the power generation enters a production plateau starting around the early June. This is due to the saturation of the 10 $KW$ \gls{PV} system.

The power ensemble range has a noticeable decrease going from May into June. This decrease correlates with the change of cloud cover at the same time. The predictability is, therefore, impacted by the cloud cover condition. The pattern can also be observed by compared the the model analysis (the red line) to the ensemble median (the black dashed line). In March when there is a significant amount of cloud cover, power simulation is less accurate, during which model analysis is barely covered by the 50\% range of the simulation ensemble. In July, on the other hand, when there is a low level of cloud cover, power simulation is more accurate and the model analysis almost overlaps with the median of the simulation ensemble. This trend is consistent also for other months of the year.

\begin{figure}
    \centering
    \includegraphics[width=\textwidth]{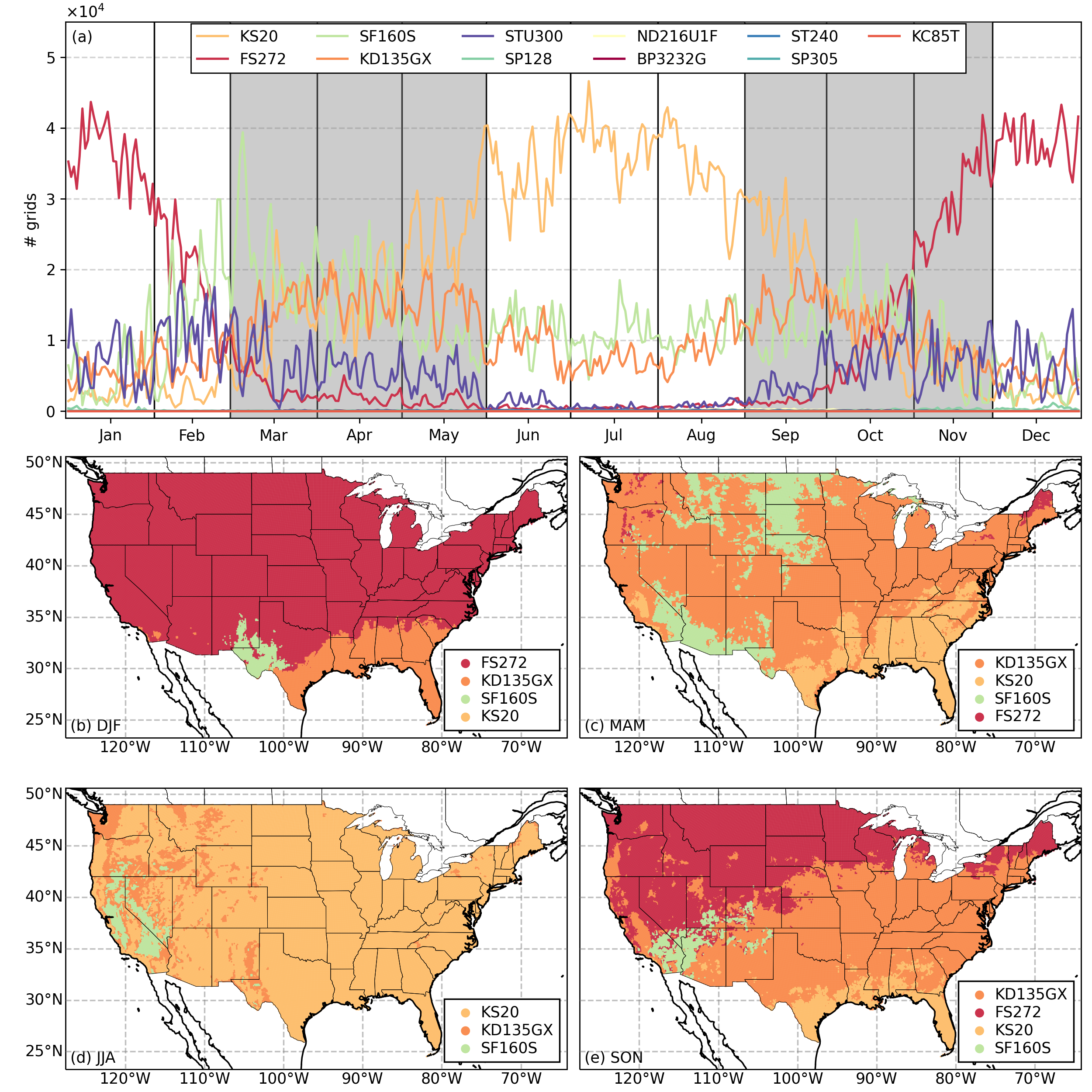}
    \caption{(a) shows the number of grid points against time where a particular \gls{PV} module has the lowest simulation \gls{RMSE}. (b, c, d, e) show the geographic map of \gls{CONUS} color coded to the corresponding \gls{PV} module with the seasonal lowest \gls{RMSE}. The labels in the legend are sorted based on the average number of grids decreasing.}
    \label{fig:best}
\end{figure}

\Fig{fig:best} compares the year-round power simulation error from 11 \gls{PV} systems both in space and time. \Fig{fig:best}a shows the total number of grid points where a particular \gls{PV} system outperforms all other modules in terms of daily \gls{RMSE} averaged across \gls{CONUS}; \Figs{fig:best}b, c, d, e, show the geographic distribution of the outperforming panel based on seasonal \gls{RMSE}.

In general, two types of modules stand out, \textit{KS20} and \textit{FS272}. These two modules have the capacity of 20 $W$ (\textit{KS20}) and 72.5 $W$ (\textit{FS272}) respectively and they are among the smallest \gls{PV} modules out of the 11 simulated modules. During the winter (\Fig{fig:best}b), \textit{FS272} is preferred on more than 85\% of the grid points. In the Northern \gls{CONUS}, \textit{FS272} is the preferred panel, while the larger panel, \textit{KD135GX}, with a capacity of 135 $W$, is preferred in the Southeastern \gls{CONUS}.    \textit{SF160S} with a capacity of 160 $W$ is preferred in Western Texas and Southern New Mexico. Since there is significantly more cloud cover over Northern \gls{CONUS} during the winter, results suggest that the power production of a \gls{PV} system with smaller module capacity but a larger module quantity can be better predicted.

During the spring when solar irradiance starts to increase, larger panels starts to take dominance (\Fig{fig:best}b). However, there might not be a clear winner because the total numbers of outperforming grids for the three panels, \textit{SF160S}, \textit{KS20}, and \textit{KD135GX}, appear to be very similar (\Fig{fig:best}a). The region that previously prefers \textit{FS272} shifts to \textit{KD135GX} with the capacity almost doubled from 72.5 $W$ to 125 $W$. This suggests the predictability of power production is related not only to weather conditions, but also how a particular \gls{PV} module responds to the changing weather conditions.

During the summer (\Figs{fig:best}d), \textit{KS20} becomes the prevailing choice thanks to its relatively small size and small capacity, therefore easy to predict. Solar irradiance typically reaches year-round climax during the summer. An exceedingly high level of irradiance leads to \gls{PV} modules generating more power than the nominal capacity but also with a higher-than-usual cell temperature. Under this weather condition, simulations have shown that smaller panels are more predictable in terms of power production. Opposite to this trend, the Pacific Northwest tends to have persistent cloud clover and relatively low solar irradiance even during the summer. Therefore, a larger panel, \textit{KD135GX}, is preferred because the working environment is not as radiant.

Finally, the fall shows a geographic transition from north to south favoring, in turns, the modules \textit{FS272}, \textit{KD135GX}, and \textit{KS20}. This is the time of the year when the solar irradiance remains relatively high while the cloud cover slowly increases. A smaller panel is preferred in the north due to high cloud cover and in the south due to over-performance.

\begin{figure}[t]
    \centering
    \includegraphics[width=\textwidth]{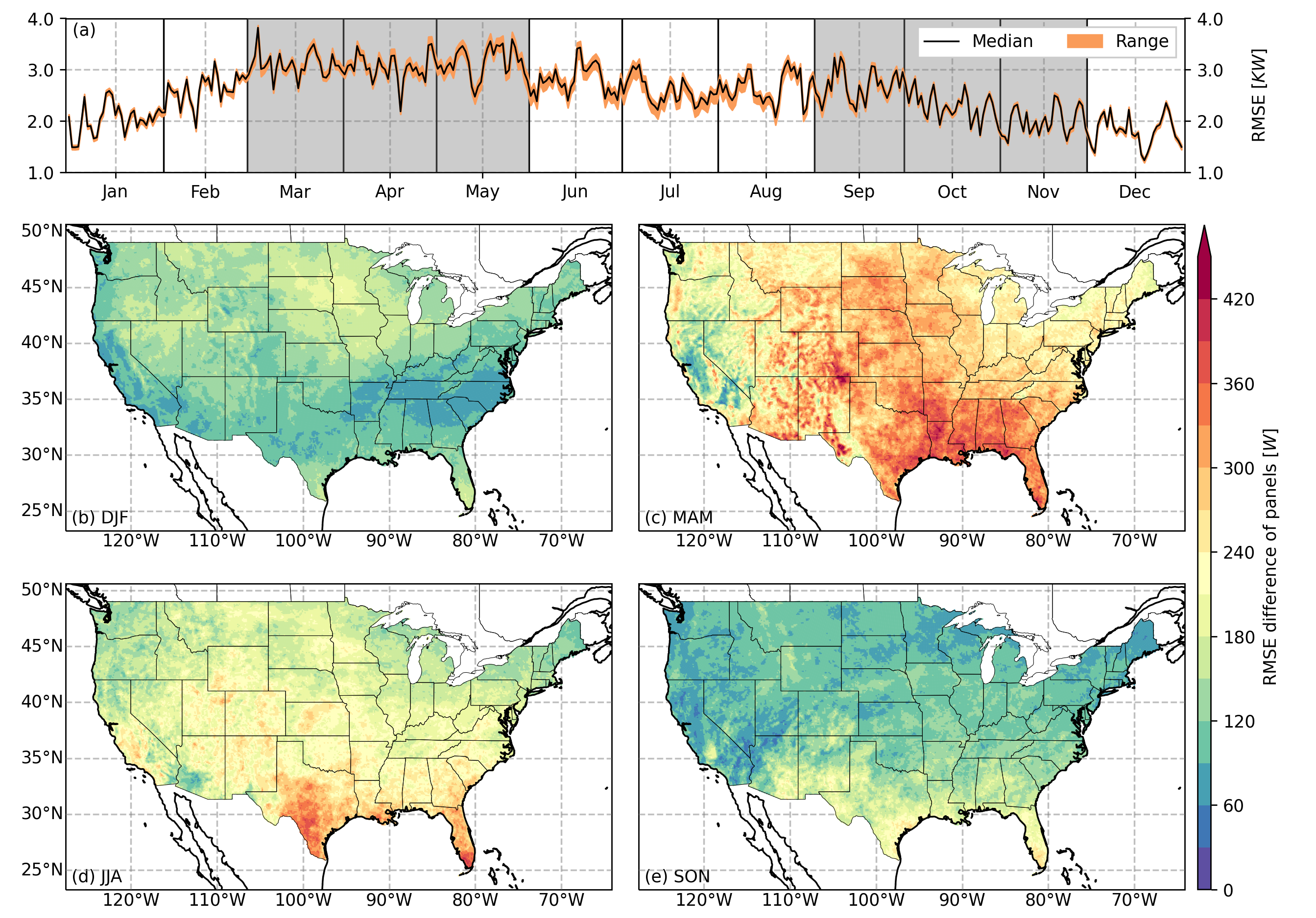}
    \caption{(a) shows the hourly power simulate error of the 10 $KW$ \gls{PV} system with the 11 simulated modules. The spread shows the range of simulate error. (b, c, d, e) show the simulate error range aggregated by seasons.}
    \label{fig:spread}
\end{figure}

\Fig{fig:best} compares the 11 \gls{PV} systems by showing outperformance, but it falls short at showing how different the power production is from the systems. \Fig{fig:spread} demonstrates how the simulation error differs among the 11 \gls{PV} systems.

\Fig{fig:spread}a shows the median of the simulation error from the 11 \gls{PV} systems in a black solid line and the range of simulation error in an orange shade. The simulation error is calculated for solar noon on each day in 2019 and then average within a particular ensemble. The range then shows the deviation of errors from 11 \gls{PV} systems.

\Figs{fig:spread}b, c, d, e show the seasonal maps of the simulation error range. For a particular grid point, the seasonal \gls{RMSE} from the worst performing module (already averaged over the 21 weather ensembles) is subtracted by the \gls{RMSE} from the best performing module. This difference in error shows the potential improvement which can be expected when using the \gls{PV} panel that is best at the location.

It is important to point out that the power simulation of a 10 $KW$ system has a very fine granularity, meaning that power prediction for each individual \gls{PV} module is linearly scaled after being simulated. The prediction error is, therefore, higher than what was shown in previous literature \cite{alessandrini_analog_2015, abuella_hourly_2017, cervone_short-term_2017, zhang_solar_2019}. The reason why the linear scaling is required is because of the unavailability of dense observed power records over \gls{CONUS}. Previous research, in fact, used observations from a limited sample of locations.  An additional difference is that the current research does not focus on predicting the output of a power plant, but rather simulating the production for individual modules, which are then scaled to the power output desired.

The \gls{RMSE} has an increasing trend in the winter and a decreasing trend in the fall (\Fig{fig:spread}a) which correlates well with the yearly pattern of solar irradiance. However, on average, the \gls{GHI} is below 500 $W/m^2$ for these two seasons (\Fig{fig:SP128}b) and \gls{PV} panels are known not to produce high power during this period. Therefore, the predictability of the 11 \gls{PV} systems is similar due to the relatively low power production, shown by the small range in \Fig{fig:spread}a and the mostly blue region in \Figs{fig:spread}b, e.

During spring and summer the predictability of power systems starts to differ. The \gls{RMSE}, on average, drops from 3 $KW$ to 2.5 $KW$ from the spring going into the summer (\Fig{fig:spread}a), due to the decrease of cloud cover. But the range of the \gls{RMSE} remains large compared to the fall and the winter. Persistent cloud cover and high level of solar irradiance in the spring are likely to affect the predictability of power production. A larger predictability difference appears in Southern and Midwestern U.S. (\Fig{fig:spread}c), reaching around 420 $W$ of \gls{RMSE} difference, equivalent to 14\% of the average \gls{RMSE} and 4.2\% of the total power generation.

During the summer, situations are slightly different because clod cover is typically low but solar irradiance is highest, reaching 870 $W/m^2$. Without the impact of cloud cover, the overall seasonal difference of predictability (\Fig{fig:spread}d) is lower compared to that of the spring. It is also shown that Southern Texas and Southern Florida have large differences in the predictability when comparing the results of the 11 \gls{PV} systems, with a \gls{RMSE} difference of over 300 $W$, equivalent to more than 12\%.

The difference in the \gls{RMSE}, or the range of the \gls{RMSE} from various \gls{PV} systems is not caused by the difference in weather conditions because all power simulations have been carried out using the same weather input generated by \gls{AnEn}. Therefore, the weather conditions are held constant when comparing power production from the \gls{PV} systems. The difference is, however, caused by how each module performs under different weather conditions. Modules have different efficiency level and are made of different materials, and therefore respond differently to weather changes. As a result, while improving weather forecasts definitely helps predicting power production, the additional uncertainty introduced by simulating the modules also needs to be studied. In fact, as discussed earlier, the characteristic of the modules can account for over 12\% of the prediction error.

\begin{table}
\caption{A summary of the year-around power simulation of all 11 \gls{PV} modules and a composite scenario where a combination of modules are used for different locations. The analysis and forecasted power generation is a summation of hourly simulation for the entire year averaged from all grid points. The \gls{RMSE} is averaged from all grid points and then normalized by the corresponding analysis column to show the percentage. Rows are sorted based on the \gls{RMSE} percentage decreasing. \text{*} indicates that the \gls{RMSE} of the composite scenario is significantly different from all other scenarios where only one module is used at the level of $p=0.05$.}
\centerline{
\begin{tabular}{crrrr}
\hline\hline
Panel & Analysis ($MW$) & Forecasted ($MW$) & RMSE ($KW$) & RMSE (\%) \\ \hline
KC85T & 25.1 & 24.49 & 745.97 & 2.97 \\
ND216U1F & 24.88 & 24.27 & 738.85 & 2.97 \\
KS20 & 24.57 & 23.98 & 730.41 & 2.97 \\
KD135GX & 24.45 & 23.86 & 722.59 & 2.95 \\
ST240 & 25.3 & 24.69 & 743.65 & 2.94 \\
STU300 & 24.83 & 24.24 & 728.84 & 2.94 \\
BP3232G & 25.5 & 24.9 & 739.08 & 2.90 \\
SP128 & 25.18 & 24.59 & 728.09 & 2.89 \\
SP305 & 25.45 & 24.86 & 731.27 & 2.87 \\
SF160S & 25.17 & 24.6 & 711.77 & 2.83 \\
FS272 & 24.88 & 24.32 & 703.02 & 2.83 \\
Composite & 24.80 & 24.24 & \textbf{699.82} & \textbf{\textit{*}2.82} \\
\hline\hline
\end{tabular}}
\label{tab:year}
\end{table}

\Table{tab:year} summaries the annual power generation and the simulation errors from the 11 simulated \gls{PV} systems from \gls{CONUS}. The power generation (the \textit{Analysis} and the \textit{Forecasted} columns) is calculated as the sum of the hourly power production for the entire year of 2019 and then averaged across \gls{CONUS} grid points. The \gls{RMSE} is the annual power production error averaged across \gls{CONUS} and then it is normalized by the analysis field to show the percentage. If two panels have the same normalized \gls{RMSE}, then they are sorted based on the \gls{RMSE}.

The largest simulated module, \textit{SP128}, has a power capacity of 400 $W$, and it is ranked \#4 (bottom-up) out of the 11 modules. The smallest simulated module, \textit{KS20}, has a power capacity of 20 $W$, and it is ranked \#9 (bottom-up) out of the modules. Although a higher power production is usually associated with a larger prediction error at the hourly temporal scale, this relation can change when analyzed at a different temporal scale, e.g., annual. On a yearly basis, \textit{SP128} is forecasted to generate 24.59 $MW$ of power, 610 $KW$ larger than that of \textit{KS20}, yet the \gls{RMSE} of \textit{SP128} still remains lower than \textit{KS20}.

\begin{figure}
    \centering
    \includegraphics[width=0.9\textwidth]{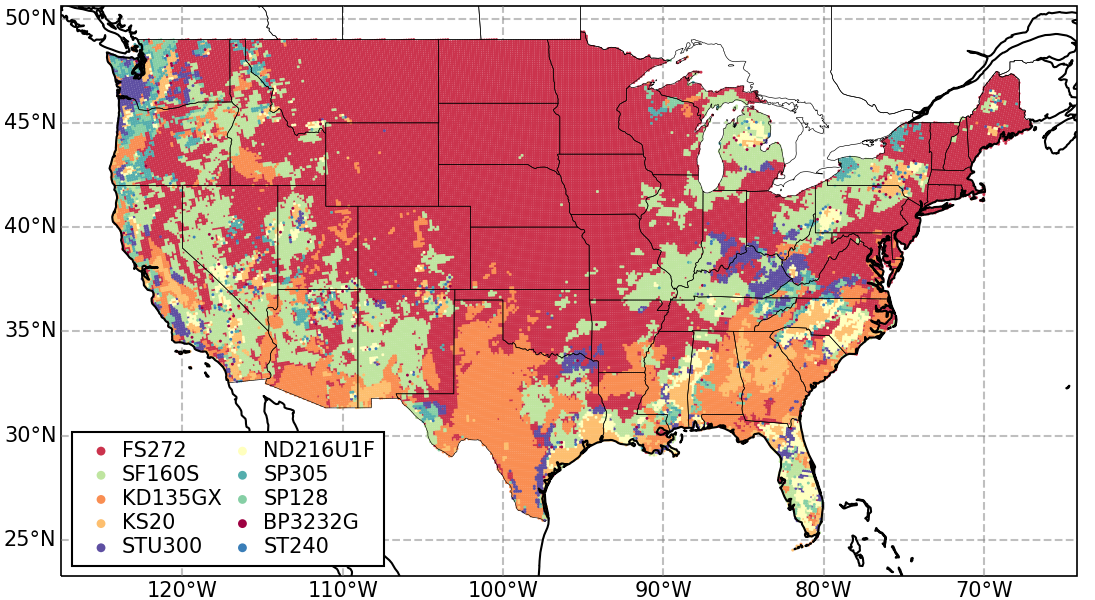}
    \caption{The map shows the geographic distribution of \gls{PV} modules used in the composite scenario. The module with the lowest \gls{RMSE} is used at each location to form the composite scenario. Coast and state boundaries are shown in solid lines. The labels in the legend are sorted column-wise based on decreasing the number grid points of a particular module.}
    \label{fig:composite}
\end{figure}

The last row of \Table{tab:year} shows the results when using the best module at each location, rather than using the same panel for the entire \gls{CONUS}. The selection of modules is based on optimizing the annual predictability and minimizing the prediction error. As a result (\Fig{fig:composite}), a majority of grid points prefer \textit{FS272} as the most predictable module, and this amounts to 31,108 grid points and 56.55\% of all grid points. Following \textit{FS272} are \textit{SF160S} (9,524 / 16.77\%), \textit{KD135GX} (7,436 / 13.10\%), \textit{KS20} (2,445 / 4.31\%), \textit{STU300} (1,728 / 3.04\%), and \textit{ND216U1F} (1,421 / 2.50\%). These mentioned panels account for more than 95\% of the grid points within \gls{CONUS} and in total, there are 10 modules selected, leaving out \textit{KC85T}. The exclusion of \textit{KC85T} is also expected from \Fig{fig:best}a that \textit{KC85T} is ranked last (bottom-up) and there are few grid points throughout the year preferring this particular module.

With the 10-module portfolio for \gls{CONUS}, the forecasted annual power generation is 24.80 $MW$ per grid and the \gls{RMSE} is 699.82 $KW$. The prediction error achieves the lowest in comparison with other single-module scenario. \Fig{fig:composite} shows the geographic distribution of the 10 modules in the composite scenario. \textit{FS272} is selected as the optimal module across the large area in the center-to-north of \gls{CONUS}. This region features a low-to-middle level solar irradiance and a middle level of total cloud cover. Since the power production is not particularly high compared to other regions of \gls{CONUS}, a smaller solar panel is preferred with 72.5 $W$ capacity. While an even smaller module, \textit{KS20}, is available, it is not chosen due to the difference in predictability. The sub-region that covers Texas and the most part of Southeast features a exceedingly high level of solar irradiance and a fair amount of cloud cover. Modules tend to perform at a higher rate than the \gls{STC} during summer. Under this condition, \textit{KD135GX} with a 135 $W$ capacity is selected for the most regions. In Florida, where a even higher level of solar irradiance is found, it is impossible to identify a single panel which consistently outperforms the others. The final pattern shown in \Fig{fig:composite} is the scattered regions favoring \textit{SF160S} which occur across different and far-away geographical areas. These regions are characterized by either low cloud cover or relatively high solar irradiance. 

\section{Enabling Scalable Simulation via the RADICAL Ensemble Toolkit}
\label{sect:entk}

RADICAL \gls{EnTK} is a workflow engine a component of the \gls{RCT} \cite{balasubramanian2019radical}. Those tools are software systems designed and implemented in accordance with the building blocks approach. Each system is independently designed with well-defined entities, functionalities, states, events and errors. Individual cybertools are designed to be consistent with a four-layered view of distributed systems for the execution of scientific workloads and workflows on \gls{HPC} resources. Each layer has a well-defined functionality and an associated ``entity''. The entities are \textbf{workflows} (or applications) at the top layer and resource specific \textbf{jobs} at the bottom layer, with \textbf{workloads} and \textbf{tasks} as intervening transitional entities in the middle layers.

\gls{RCT} has three main components: \gls{RS} \cite{merzky2015saga}, \gls{RP} \cite{merzky2021design,merzky2018using} and RADICAL \gls{EnTK} \cite{balasubramanian2018harnessing,balasubramanian2016ensemble}. \gls{RS} is a Python implementation of the Open Grid Forum SAGA standard GFD.90 \cite{goodale2006saga}, a high-level interface to distributed infrastructure components like job schedulers, file transfer and resource provisioning services. \gls{RS} enables interoperability across heterogeneous distributed infrastructures, improving on their usability and enhancing the sustainability of services and tools. \gls{RP} is a Python implementation of the pilot paradigm and architectural pattern~\cite{turilli2018comprehensive}. Pilot systems enable users to submit pilot jobs to computing infrastructures and then use the resources acquired by the pilot to execute one or more tasks. These tasks are directly scheduled via the pilot, without having to queue in the infrastructure's batch system.

\gls{EnTK} is a Python implementation of a workflow engine, specialized in supporting the programming and execution of applications with ensembles of tasks. \gls{EnTK}  executes tasks concurrently or sequentially, depending on their arbitrary priority relation. Tasks are scalar, MPI, OpenMP, multi-process, and multi-threaded programs that run as self-contained executables. Tasks are not
functions, methods, threads or subprocesses.

\subsection{Application Model}

\gls{EnTK} supports \gls{EBA}, where an ensemble is defined as a set of tasks. Ensembles may vary in the number of tasks, types of task, tasks' executable (or executing kernel), and tasks' resource requirements. \gls{EBA} may vary in the number of ensembles or the runtime dependencies among ensembles. The space of \gls{EBA} is vast, and thus there is a need for simple and uniform abstractions while avoiding single- point solutions.

Currently, the use cases motivating \gls{EnTK} require a maximum of \(O(10^4)\) tasks but \gls{EnTK} is designed to support up to \(O(10^6)\) tasks due to the rate at which this requirement has been increasing, especially in biomolecular simulations where a greater number of tasks is associated with greater sampling or more precise free energy calculations. \gls{EnTK} supports resubmission of failed tasks, without application checkpointing, and restarting of failed RTS and components. In this way, applications can be executed on multiple attempts, without restarting completed tasks.

\gls{EnTK} models \gls{EBA} by combining the following user-facing constructs:

\begin{itemize}
  \item \textbf{Task:} an abstraction of a computational task that contains
  information regarding an executable, its software environment and its data
  dependences.
  \item \textbf{Stage:} a set of tasks without mutual dependences and that
  can be executed concurrently.
  \item \textbf{Pipeline:} a list of stages where any stage \(i\) can be
  executed only after stage \(i-1\) has been executed.
\end{itemize}

The application consists of a set of pipelines, where each pipeline is a list of stages, and each stage is a set of tasks. All the pipelines can execute concurrently, all the stages of each pipeline can execute sequentially, and all the tasks of each stage can execute concurrently. Pipelines, stages, and tasks (PST) descriptions can be extended to account for dependencies among groups of pipelines in terms of lists of sets of pipelines. Further, the specification of branches in the execution flow of applications does not require to alter the PST semantics: Branching events can be specified as tasks where a decision is made about the runtime flow. For example, a task could be used to decide to skip some elements of a stage, based on some partial results of the ongoing computation.

\subsection{Architecture} 

\begin{figure}
 \includegraphics[width=\columnwidth]{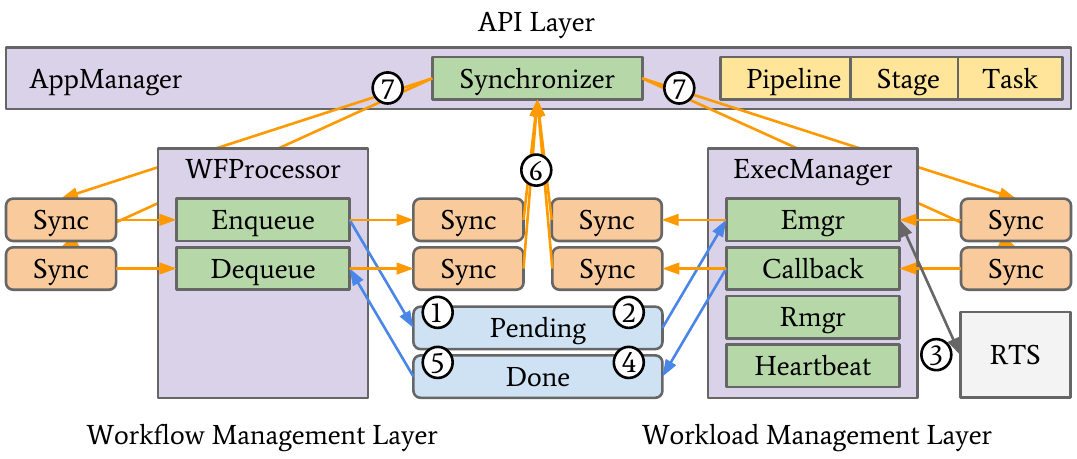}
 \caption{\gls{EnTK} architecture and execution model. Components' (purple) subcomponents (green) use  queues (blue and orange) to communicate and coordinate the execution of an application via a chosen RTS (gray).}
 \label{fig:entk-arch}
\end{figure}

\gls{EnTK} sits between the user and the \gls{HPC}, abstracting resource management and execution management from the user. \Fig{fig:entk-arch} shows the components (purple) and subcomponents (green) of \gls{EnTK}, organized in three layers: \gls{API}, Workflow Management, and Workload Management.

The \gls{API} layer enables users to codify PST descriptions. The Workflow Management layer retrieves information from the user about available infrastructures, initializes \gls{EnTK}, and holds the global state of the application during execution. The Workload Management layer acquires resources via the RTS. The Workflow Management layer has two components: AppManager and WFProcessor. AppManager uses the Synchronizer subcomponent to update the state of the application at runtime. WFProcessor uses the Enqueue and Dequeue subcomponents to queue and dequeue tasks from the Workload Management layer. The Workload Management layer uses ExecManager and its Rmgr, Emgr, RTS Callback, and Heartbeat subcomponents to acquire resources from infrastructures and execute the application. Another benefit of this architecture is the isolation of the RTS into a stand-alone subsystem. This enables composability of \gls{EnTK} with diverse RTS and, depending on capabilities, multiple types of infrastructures. Further, \gls{EnTK} assumes the RTS to be a black box enabling fault-tolerance. When the RTS fails or becomes unresponsive, \gls{EnTK} can tear it down and bring it back, loosing only those tasks that were in execution at the time of the RTS failure.

\subsection{Execution Model}

\gls{EnTK} components and subcomponents communicate and coordinate for the execution of tasks. Users describe an application via the \gls{API}, instantiate the AppManager component with information about the available infrastructures and then pass the application description to AppManager for execution. AppManager holds these descriptions and, upon initialization, creates all the queues, spawns the Synchronizer, and instantiates the WFProcessor and ExecManager. WFProcessor and ExecManager instantiate their own subcomponents.

Once \gls{EnTK} is fully initialized, WFProcessor initiates the execution by creating a local copy of the application description from AppManager and tagging tasks for execution. Enqueue pushes these tasks to the Pending queue (\textcircled{\raisebox{-0.96pt}1} in \Fig{fig:entk-arch}). Emgr pulls tasks from the Pending queue (\textcircled{\raisebox{-0.96pt}2} in \Fig{fig:entk-arch}) and executes them using a RTS (\textcircled{\raisebox{-0.96pt}3} in \Fig{fig:entk-arch}). RTS Callback pushes tasks that have completed execution to the Done queue (\textcircled{\raisebox{-0.96pt}4} in \Fig{fig:entk-arch}). Dequeue pulls completed tasks (\textcircled{\raisebox{-0.96pt}5} in \Fig{fig:entk-arch}) and tags them as done, failed or canceled, depending on the return code from the RTS\@.

Throughout the execution of the application, tasks, stages and pipelines undergo multiple state transitions in both WFProcessor and ExecManager. Each component and subcomponent synchronizes these transitions with AppManager by pushing messages through dedicated queues (\textcircled{\raisebox{-0.96pt}6} in \Fig{fig:entk-arch}).AppManager pulls these messages and updates the application states. AppManager then acknowledges the updates via dedicated queues (\textcircled{\raisebox{-0.96pt}7} in \Fig{fig:entk-arch}). This messaging mechanism ensures that AppManager is always up-to-date with any state change, making it the only stateful component of EnTK\@.

\subsection{Integration of Parallel Analog Ensemble}

The integration of the \gls{PAnEn} and the RADICAL \gls{EnTK} provides good examples for the ensemble-of-pipelines model and the pipeline-of-ensembles model.

The ensemble-of-pipelines model refers to the workflow where each pipeline is an independent execution unit that might consist of several stages and there are several such pipelines to execute in parallel. This type of model is adopted for weight optimization shown in \Sect{sect:weight}. In total, there are 8,001 different weight combinations to experiment and each trial is independent from each other, meaning that the execution with one weight combination does not need to wait for or communicate with any other executions. Each trial consists of a pipeline following the process of generating weather analogs, simulation power production, and finally verifying power simulation results with the forecast analysis. The three stages should be executed in series and therefore, they have been placed with order in a pipeline. Since the processes need to be repeated for the \gls{NN} and the \gls{RB}, the final individual pipeline has six tasks in total.

The pipeline-of-ensembles model refers to the workflow where there are multiple sequential stages and each stage consists of tasks that will be executed in parallel. The later simulation of the 11 \gls{PV} modules adopted this model by two stages:

\begin{enumerate}
    \item Stage one: \gls{AnEn} forecasts have been generated for each of the subset domain in parallel.
    \item Stage two: Each of the 11 modules is used to simulate power production for the corresponding subset domain.
\end{enumerate}

Note that since the computation for each subset domain is considered as an individual task, tasks can be executed in parallel. There are 71 subset domains for the \gls{NN} and 28 for the \gls{RB}. On the other hand, stages are executed in sequence because the power simulation can only be carried out after the \gls{AnEn} forecasts have been successfully generated. \gls{EnTK} offers a flexible interface for defining the computational requirement for each task so that each task can dynamically request a proper amount of allocation based on the area of the subset domain. Given that there are dozens of subsets with different area, this functionality frees the developer from having to specify the computation manually.

\begin{table}
\caption{Summary of experiment runtime and allocation request on \gls{NCAR} Cheyenne}
\centerline{
\scalebox{0.88}{
\begin{tabular}{ccrr}
\hline\hline
Experiment & Size of workflow & \# cores & Runtime (h) \\ \hline
\gls{AnEn} weight optimization & 8,001 pipelines x 6 tasks per pipeline & 7,200 & 1.31\\
Ensemble power simulation & 2 stages x 99 tasks per stage & 5,508 & 4.28\\
\hline\hline
\end{tabular}}}
\label{tab:runtime}
\end{table}

In general, the pipeline-of-ensembles model offers more efficient concurrency because it utilizes lightweight tasks. As a result, it is recommended to adopt this type of model as much as possible. The ensemble-of-pipelines model merely provides another layer of abstraction for users to create more complex scientific workflows. \Table{tab:runtime} summarizes the runtime and computational requirement of the two experiments executed on \gls{NCAR} Cheyenne.

\section{Final Remarks}
\label{sect:final}

Forecast ensembles typically offer higher prediction accuracy than deterministic predictions, and they provide additional information on uncertainty. However, running a ensemble \gls{NWP} model usually involves both technical and computational challenges. Researchers often download forecast ensembles from archives, but the amount of accessed data can easily exceed dozens of gigabyte, sometimes terabyte. Designing a scientific workflow that involves simulation ensembles require parallelization and scalable tools.

This chapter demonstrates the ability of using the \gls{AnEn} method and the RADICAL \gls{EnTK} tools to design ensemble workflows at scale. First, deterministic predictions over \gls{CONUS} for 2019 has been transformed into 21-member forecast ensembles using \gls{AnEn}. The forecasts have a spatial resolution of 12 km and a temporal resolution of one hour. Second, the weather forecast ensembles are used to simulate point-by-point power production with 11 different \gls{PV} modules. Over 110 TB of simulation data have been generated on the \gls{NCAR} supercomputer, Cheyenne.

Results have shown that the predictability of solar power generation depends on not only the meteorological variation, but also the configuration of \gls{PV} modules. In order to optimize the amount of predictable solar power, the type of \gls{PV} modules should be an important factor to consider in its relation to geographic locations and its interplay with long-term weather conditions.


\appendix
\section{A Minimal Workflow for Power Simulation}
\label{app:core}

\begin{lstlisting}[language=Python, caption={An example of \gls{PV} power simulation with \textit{PV\_LIB} in Python}]
import pandas as pd
from pvlib import pvsystem, irradiance, location, atmosphere, temperature

def simulate_power_production(
        timestamp, latitude, longitude, ghi, alb, tamb, wspd,
        tc_module, pv_module):
    """
    The core function for simulating power production under a given forecasted
    weather condition and a solar panel configuration at a particular location.

    This function is developed and tested with pvlib v0.8.1.
    """

    # Initialize a location
    loc = location.Location(latitude=latitude, longitude=longitude)

    # Calculate the solar position
    solar_position = loc.get_solarposition(timestamp)

    # Calculate extraterrestrial irradiance
    dni_extra = irradiance.get_extra_radiation(timestamp)

    # Calculate air mass
    air_mass = atmosphere.get_relative_airmass(solar_position['apparent_zenith'])

    # Decompose DNI from GHI
    dni_dict = irradiance.disc(ghi, solar_position['zenith'], timestamp)

    # Calculate sky diffuse
    sky_diffuse = irradiance.haydavies(
        0, 180, ghi, dni_dict['dni'], dni_extra,
        solar_position['apparent_zenith'], solar_position['azimuth'])

    # Calculate ground diffuse
    ground_diffuse = irradiance.get_ground_diffuse(0, ghi, alb)

    # Calculate angle of incidence
    aoi = irradiance.aoi(0, 180, solar_position['apparent_zenith'],
                         solar_position['azimuth'])

    # Calculate in-plane irradiance components
    poa = irradiance.poa_components(
        aoi, dni_dict['dni'], sky_diffuse, ground_diffuse)

    # Calculate cell temperature
    tcell = pvsystem.temperature.sapm_cell(
        poa['poa_global'], tamb, wspd, **tc_module)

    # Calculate effective irradiance
    effective_irradiance = pvsystem.sapm_effective_irradiance(
        poa['poa_direct'], poa['poa_diffuse'], air_mass, aoi, pv_module)

    # Calculate power
    sapm_out = pvsystem.sapm(effective_irradiance, tcell, pv_module)

    return sapm_out

# Example usage
pv_module = pvsystem.retrieve_sam('SandiaMod')['Silevo_Triex_U300_Black__2014_']
tc_module = temperature.TEMPERATURE_MODEL_PARAMETERS[
    'sapm']['open_rack_glass_polymer']
timestamp = pd.to_datetime('2021-01-01 12:00:00')

simulate_power_production(timestamp, -77.86, 40.79, 400, 0.8, 27, 1.2,
                          tc_module, pv_module)
\end{lstlisting}

\end{document}